\documentclass[aps,prd,superscriptaddress,showkeys,showpacs,a4paper,nofootinbib,twocolumn]{revtex4-2}

\usepackage[english]{babel}
\usepackage{amsmath,bm}
\usepackage{amssymb}
\usepackage{enumitem}
\usepackage{textcomp}
\usepackage{graphicx,float}
\usepackage{dsfont}
\usepackage{color}

\usepackage[breaklinks=true]{hyperref}
\hypersetup{
	colorlinks=true,
	linkcolor=blue,
	filecolor=magenta,      
	urlcolor=blue,
	citecolor=red,
}

\usepackage{mathrsfs}
\usepackage[normalem]{ulem}

\usepackage{placeins}

\definecolor{dgreen}{rgb}{0.,0.5,0}

\begin{document}

{\hfill MS-TP-23-06}
	
\title{Backreaction from gauge fields produced during inflation}

\author{R.~Durrer}
\affiliation{D\'{e}partement de Physique Th\'{e}orique and Center for Astroparticle Physics, Universit\'{e} de Gen\`{e}ve,  24 quai Ernest Ansermet, 1211 Gen\`{e}ve 4, Switzerland}

\author{O.~Sobol}
\email{oleksandr.sobol@knu.ua}
\affiliation{Institute for Theoretical Physics, University of M\"{u}nster, 9, Wilhelm-Klemm-Stra{\ss}e, 48149 M\"{u}nster, Germany}
\affiliation{Physics Faculty, Taras Shevchenko National University of Kyiv, 64/13, Volodymyrska Street, 01601 Kyiv, Ukraine}
	
\author{S.~Vilchinskii}
\affiliation{D\'{e}partement de Physique Th\'{e}orique and Center for Astroparticle Physics, Universit\'{e} de Gen\`{e}ve,  24 quai Ernest Ansermet, 1211 Gen\`{e}ve 4, Switzerland}
\affiliation{Physics Faculty, Taras Shevchenko National University of Kyiv, 64/13, Volodymyrska Street, 01601 Kyiv, Ukraine}

\date{\today}
	
\begin{abstract}
In this work, we study general features of a regime where gauge fields produced during inflation cause a strong backreaction on the background evolution and its impact on the spectrum and the correlation length of gauge fields.
With this aim, the gradient-expansion formalism previously proposed for the description of inflationary magnetogenesis in purely kinetic or purely axial coupling models, is  extended  to the case when both types of coupling are present. As it is formulated in position space, this method allows us to self-consistently take into account the backreaction of generated gauge fields on the inflationary background because it captures the nonlinear evolution of all physically relevant gauge-field modes at once. 
Using this extended gradient-expansion formalism, suitable for a wide range of inflationary magnetogenesis models, we study the gauge-field production in a specific generalization of the Starobinsky $R^2$-model with a nonminimal coupling of gauge fields to gravity. In the Einstein frame, this model implies, in addition to an asymptotically flat inflaton potential, also a nontrivial form of kinetic and axial coupling functions which decrease in time and, thus, are potentially suitable for the generation of  gauge fields with a scale-invariant or even red-tilted power spectrum. 
The numerical analysis shows, however, that  backreaction, which unavoidably occurs in this model for the interesting range of parameters, strongly alters the behavior of the spectrum and does not allow us to obtain a sufficiently large correlation length for the magnetic field. The oscillatory behavior of the generated field, caused by the retarded response of the gauge field to changes of the inflaton velocity, was  revealed.
\end{abstract}

\maketitle

\section{Introduction}
\label{sec-intro}

The paradigm of  inflation~\cite{Starobinsky:1980,Guth:1981,Linde:1982,Starobinsky:1982,Albrecht:1982,Linde:1983} is a very successful idea which not only solves a large number of cosmological problems, but also provides a convincing mechanism for the origin of the primordial fluctuations (the seeds of the galaxies and of the cosmic microwave background (CMB) anisotropies)~\cite{Starobinsky:1979,Mukhanov:1981,Mukhanov:1982,Guth:1982,Hawking:1982,Bardeen:1983} and predicts that their spectrum should be almost scale invariant.
At present, some inflationary models are in good agreement with observations \cite{Planck:2013-infl,Planck:2015-infl,Planck:2018-infl} (for a review, see Refs.~\cite{Martin:2018,Martin:2014}).
The majority of the mechanisms, which provide a period of inflation, can be regarded as classical. 
Nevertheless, the quantum nature of the underlying fundamental physics always leaves its imprints on  cosmological observables. 
The first imprint are the temperature anisotropies and  polarization of CMB. The properties of these anisotropies provide a direct link to inflation. Apart from the CMB, any signal with an extragalactic coherence length, of the order of Megaparsec (Mpc) and larger, also can be treated as a candidate for  an inflationary imprint. Since scales larger than a Mpc were outside the causal horizon before the epoch of structure formation, it is very difficult to find causal mechanisms capable of generating correlations on these scales. A very important example of such a signal are the observed extragalactic magnetic fields with very large coherence scales of $\lambda_{B}\gtrsim 1\,$Mpc and with strength $10^{-16}\lesssim B_{0}\lesssim 10^{-10}\,$Gauss.
These magnetic fields in voids were indirectly detected recently through the  gamma-ray observations of distant blazars~\cite{Tavecchio:2010,Ando:2010,Neronov:2010,Dolag:2010,Dermer:2011,Taylor:2011,Caprini:2015,MAGIC:2022}; their magnitude is also constrained from observations of the CMB \cite{Planck:2015-pmf,Sutton:2017,Giovannini:2018b,Paoletti:2018,Brandenburg:2020vwp} and from ultra-high-energy cosmic rays~\cite{Bray:2018,Neronov:2021}.
Large coherence length, measured in Mpc, suggests that these magnetic fields may have been generated during the earliest stage of the Universe evolution, i.e., during inflation (see, e.g., Ref.~\cite{Durrer:2013}). If  magnetic fields observed in voids are really of primordial origin, they represent a new source of information whose statistics will contain information about gauge fields in the very early Universe.
 
In order to produce gauge fields during inflation, the conformal invariance of the corresponding Maxwell action has to be broken, as fluctuations of a massless gauge field are not generated in a conformally flat inflationary background \cite{Parker:1968}. Conformal invariance can be broken, e.g., by  coupling gauge fields to the scalar or pseudoscalar inflaton field or to spacetime curvature, see the seminal works \cite{Turner:1988,Ratra:1992,Garretson:1992,Dolgov:1993}, which were revisited and improved later many times~\cite{Giovannini:2001,Bamba:2004,Martin:2008,Kanno:2009,Demozzi:2009,Maleknejad:2012,Ferreira:2013,Ferreira:2014,Vilchinskii:2017,Sharma:2017b,Sobol:2018,Talebian:2020,Sobol:2020lec,Sasaki:2022,Durrer:2011,Anber:2006,Anber:2010,Barnaby:2012,Caprini:2014,Anber:2015,Ng:2015,Fujita:2015,Adshead:2015,Adshead:2016,Notari:2016,Domcke:2018eki,Cuissa:2018,Shtanov:2019,Sobol:2019,Domcke:2020,Caravano:2021,Gorbar:2021rlt,Gorbar:2022,Adshead:2018doq,Adshead:2019igv,Adshead:2019lbr,Caravano:2022,Bastero-Gil:2022,Fujita:2022,Savchenko:2018,Sobol:2021,Durrer:2022emo,Maity:2021,Bamba:2008,Bamba:2020,Cecchini:2023}. Two models of magnetogenesis have receiver most attention, namely the kinetic~\cite{Giovannini:2001,Bamba:2004,Martin:2008,Kanno:2009,Demozzi:2009,Maleknejad:2012,Ferreira:2013,Ferreira:2014,Vilchinskii:2017,Sharma:2017b,Sobol:2018,Talebian:2020,Sobol:2020lec,Sasaki:2022} and axial~\cite{Durrer:2011,Anber:2006,Anber:2010,Barnaby:2012,Caprini:2014,Anber:2015,Ng:2015,Fujita:2015,Adshead:2015,Adshead:2016,Notari:2016,Domcke:2018eki,Cuissa:2018,Shtanov:2019,Sobol:2019,Domcke:2020,Caravano:2021,Gorbar:2021rlt,Gorbar:2022,Adshead:2018doq,Adshead:2019igv,Adshead:2019lbr,Caravano:2022,Bastero-Gil:2022,Fujita:2022} coupling models which are described by terms in the Lagrangians $\propto I_1(\phi)F_{\mu\nu}F^{\mu\nu}$ and $\propto I_2(\phi)F_{\mu\nu}\tilde{F}^{\mu\nu}$ respectively, where $\tilde{F}^{\mu\nu}$ is the dual gauge-field tensor. Different types of gauge-field couplings to the curvature scalar and/or tensors during slow-roll inflation effectively boil down to the above-mentioned models (up to corrections suppressed by the slow-roll parameters)~\cite{Savchenko:2018,Sobol:2021,Durrer:2022emo}. This is because the expansion of the produced is approximately described by the de Sitter solution which represents a maximally symmetric spacetime, is fully characterized by the curvature scalar which can be expressed as a function of the inflaton field. Thus, a general model with both kinetic and axial couplings, as considered in the present article, covers almost all possible magnetogenesis models during slow-roll inflation except maybe some exotic scenarios.

It should be noted that any model of magnetogenesis typically faces a backreaction problem during inflation, wherein the produced gauge field impacts the time evolution of the inflaton field and the Universe expansion rate. This may lead to a prolongation of the inflation stage, have a strong impact on the spectra of primordial perturbations, and, moreover, can strongly modify the reheating stage~\cite{Kanno:2009,Ferreira:2014,Sobol:2018,Barnaby:2012,Sobol:2019,Domcke:2020,Gorbar:2021rlt,Adshead:2018doq,Adshead:2019igv,Adshead:2019lbr,Caravano:2022,Bastero-Gil:2022}. Therefore, it is very important to keep backreaction under control because it may spoil the predictions of a given inflationary model.

However, this appears to be a very nontrivial task. Indeed, in the absence of backreaction, it is possible first to solve equations governing the evolution of the inflaton field and the scale factor without gauge fields and then describe the gauge-field production on such a predetermined background. Then, the equations of motion for the gauge field remain linear and the most straightforward way to describe magnetogenesis is to track the evolution of separate Fourier modes in momentum space. However, this approach is applicable only if the generated fields are weak enough not to affect the background evolution. Otherwise, one has to take into account the generated gauge fields nonperturbatively and solve all equations of motion simultaneously. Clearly, this makes the problem of inflationary magnetogenesis much harder because the joint dynamics of the inflaton and gauge fields now becomes highly nonlinear.

In order to take into account backreaction one can follow different strategies. First, it is still possible to work in momentum space and track the evolution of some finite number of Fourier modes simultaneously. However, this may be a nontrivial task since the range of modes crossing the horizon during inflation spans over many orders of magnitude. Second, one can separate the gauge-field dynamics from the background by employing an iterative approach: the gauge field on the $n$th iteration is determined from the background on the $(n-1)$th iteration. This approach has been applied to describe magnetogenesis from axion inflation in Ref.~\cite{Domcke:2020}. Alternatively, one can switch to position space and consider all interacting fields on a grid where the Universe expansion is incorporated~\cite{Cuissa:2018,Caravano:2021}. In this work we would like to consider yet another approach which is also working in position space but operates with quantum average quantities which apparently are independent of spatial coordinates. This is the gradient expansion formalism\footnote{We would like to note that the term ``gradient expansion'' is widely used in the literature to denote the approximate method in which one performs an expansion (of action or EOMs) in powers of spatial derivatives provided that the field slowly changes in space. In particular, this technique is often used in cosmological perturbation theory for the description of modes well beyond the Hubble horizon, see, e.g., Ref.~\cite{Sugiyama:2012}. However, in this work, the term gradient expansion does not mean any approximation or expansion into power series; it is related to the fact that our method is based on an infinite set of bilinear gauge-field functions with increasing number of spatial derivatives (gradients).} which was proposed for the description of inflationary magnetogenesis in our papers~\cite{Sobol:2020lec} for the purely kinetic coupling model and in~\cite{Sobol:2019,Gorbar:2021rlt} for the case of axial coupling. This method is based on a system of ordinary differential equations for a set of observables given by vacuum expectation values of scalar products of electric and/or magnetic fields with an arbitrary number of spatial curls acting on them. This system, truncated at a certain order (maximal number of curls), allows to describe the generated gauge field with an accuracy typically of order a few percent. In the present work we generalize this formalism to the case where both types of coupling are present.

In order to capture only those gauge-field modes which are enhanced due to kinetic and axial couplings to the inflaton field, we introduce an ultraviolet cutoff in momentum\,---\,the so called horizon-crossing momentum $k_\mathrm{h}$\,---\,and include in the bilinear gauge-field functions only contributions of modes with momenta $k\leq k_\mathrm{h}$. The threshold momentum $k_\mathrm{h}$ grows in time during inflation which means that more and more new modes make their contributions to the bilinear functions as the time passes. This additional time dependence of the bilinear functions is described by the boundary terms in their equations of motion which act like vacuum sources. Another example where short-wavelength modes play the role of a source for the dynamics of long-range modes is given in a recent article~\cite{Fujita:2022}, where the authors propose a stochastic approach to inflationary magnetogenesis and work directly with the vectors of electric and magnetic field. They obtain a system of equations for the infrared modes with a stochastic noise originating from the ultraviolet ones and solve it numerically in order to get a picture of stochastic behavior of the produced gauge field. Although the idea of the mode separation is rather similar to ours, there are two key differences compared to our work: (i)~the threshold momentum in Ref.~\cite{Fujita:2022} is much smaller than our value of $k_\mathrm{h}$, because they want both infrared and ultraviolet modes to behave classically; (ii)~they study the stochastic evolution of gauge fields while we consider only deterministic functions\,---\,variances of gauge-field vectors, i.e., we perform the statistical averaging at the very beginning when we define the dynamical variables. In a certain sense, our boundary terms may be regarded as analogs of the two-point correlation functions of the stochastic noise. Thus, the stochastic approach of Ref.~\cite{Fujita:2022} allows to look at inflationary magnetogenesis from a different perspective.

Another important question is the choice of a model of inflationary magnetogenesis. The arbitrariness of the kinetic and axial coupling functions, on one hand, gives one the possibility to construct a model explaining potentially any observation, while on the other hand, it reduces the predictive power of such a theory. In some sense, this situation is analogous to the problem of fixing the inflationary model. Potentially, there is an infinite number of different inflaton potentials if they are constructed phenomenologically (a large but not comprehensive list of inflationary models can be found, e.g., in Ref.~\cite{Martin:2014}). This is often considered a drawback of the inflationary paradigm. However, one can think of an economical way to provide inflation without introducing any extra fields and any ``manual'' construction of the potential. For instance, the Starobinsky $R^2$-model~\cite{Starobinsky:1980} extends a gravity sector by a term $\propto R^2$ (which would anyway appear in a quantum theory of gravity due to quantum corrections~\cite{Fischetti:1979}); remarkably, in the Einstein frame, this theory reveals an additional scalar degree of freedom whose potential has an asymptotically flat form which is favored by CMB observations~\cite{Planck:2018-infl}. Another example is Higgs inflation~\cite{Bezrukov:2008,Bauer:2008} where the Standard Model Higgs field supplied by a nonminimal coupling to curvature plays the role of the inflaton, again with a perfectly suitable potential. A combination of two previous models\,---\,the Higgs-Starobinsky model~\cite{Ema:2017,Starobinsky:2018,Gorbunov:2018}\,---\,is also in accordance with the CMB observations and provides a ultraviolet completion for Higgs inflation.

A similar economical way of magnetogenesis model building also looks more convincing. For example, one can nonminimally couple the gauge field to a spacetime curvature which  fixes the coupling functions up to a finite set of constant parameters. This idea was realized in the context of Starobinsky inflation~\cite{Savchenko:2018}, general $f(R)$-inflation~\cite{Bamba:2008}, Higgs inflation~\cite{Sobol:2021}, and recently in the Higgs-Starobinsky model~\cite{Durrer:2022emo}. A typical problem which arises in all three cases is that, in the Einstein frame, the corresponding action potentially contains higher powers of the gauge field which make the theory nonlinear and ultraviolet-incomplete. Therefore, only a perturbative treatment of magnetogenesis is possible.

For this reason, in this work, we propose an extension of the Starobinsky model by promoting the coefficient in front of the $R^2$-term to a function of the gauge field. The corresponding term in the Lagrangian reads as
$$
\Delta\mathcal{L}= \frac{\xi_s}{4} \Big[1+\frac{\kappa_{1}}{M_{\mathrm{P}}^4}F_{\mu\nu}F^{\mu\nu}+\frac{\kappa_{2}}{M_{\mathrm{P}}^4} F_{\mu\nu}\tilde{F}^{\mu\nu}\Big]^{-1} R^2,
$$
where $\xi_s$, $\kappa_1$, and $\kappa_2$ are the constant dimensionless parameters. The explicit form of this function is chosen such that in the Einstein frame the following conditions are fulfilled: (i)~the inflaton potential remains unchanged, (ii)~the problem avoids strong coupling in the gauge sector. (iii)~the action remains quadratic in the gauge field. The latter condition implies that the gauge-field generation may be studied nonperturbatively, including  backreaction, by means of the gradient-expansion formalism.
 
The work presented in this paper is organized as follows. In Sec.~\ref{sec-basics}, we 
present a general model of inflationary magnetogenesis with kinetic and axial couplings between an Abelian gauge field and the inflaton. Section~\ref{sec-gef} is devoted to the derivation of the gradient-expansion formalism in the case when both kinetic and axial couplings are present. Here we introduce the set of observables, derive equations of motion for them, discuss the origin of boundary terms and find their explicit expressions, and describe how to truncate the series of equations at some finite order. In Sec.~\ref{sec-nonminimal}, we propose an extension of the Starobinsky inflationary model which includes interaction with the gauge field and we rewrite its action in the Einstein frame. It takes precisely the form of the gauge-field action with kinetic and axial coupling to the inflaton, where the corresponding coupling functions are fixed up to constant parameters. In Sec.~\ref{sec-numerical}, we then study the gauge-field production in this model numerically using the gradient-expansion formalism. We explore different cases of kinetic-dominated, axial-dominated, and mixed coupling, and we investigate the general features of backreaction. Section~\ref{concl} is devoted to conclusions. 
In Appendix~\ref{app-Whittaker}, we discuss the Whittaker equation and properties of its solutions which describe the time evolution of gauge-field mode functions. In Appendix~\ref{app-B}, we provide more details on our numerical procedure used to solve the system of equations of the gradient-expansion formalism.

Throughout the work we use the natural units and set $\hbar=c=1$; we use the notation $M_{\mathrm{P}}=(8\pi G)^{-1/2}\approx 2.43\times 10^{18}\,{\rm GeV}$ for the reduced Planck mass.

\section{Magnetogenesis from general kinetic and axial couplings}
\label{sec-basics}

Let us consider an Abelian gauge field $A_{\mu}$ which interacts with the inflaton field $\phi$ via kinetic and axial couplings. The corresponding action reads as
\begin{multline}
\label{action-kinetic-axial}
S_{\mathrm{KA}}[g_{\mu\nu},\phi, A_{\mu}]=\int d^4 x \sqrt{-g}\Big[\frac{1}{2}g^{\mu\nu}\partial_\mu \phi\partial_\nu \phi-V(\phi)\\
-\frac{1}{4} I_1(\phi)F_{\mu\nu}F^{\mu\nu}-\frac{1}{4}I_2(\phi)F_{\mu\nu}\tilde{F}^{\mu\nu}\Big],
\end{multline}
where $g_{\mu\nu}$ is the spacetime metric, $g=\mathrm{det}\,g_{\mu\nu}$, $V(\phi)$ is the inflaton potential, $F_{\mu\nu}=\partial_{\mu}A_{\nu}-\partial_{\nu}A_{\mu}$ is the gauge-field tensor, $\tilde{F}^{\mu\nu}=
\frac{\varepsilon^{\mu\nu\alpha\beta}}{2\sqrt{-g}} F_{\alpha\beta}$ is its dual, where $\varepsilon^{\mu\nu\alpha\beta}$ is the absolutely antisymmetric Levi-Civita symbol with $\varepsilon^{0123}=+1$. The function $I_1(\phi)$ [$I_2(\phi)$] describes the kinetic [axial] coupling of the gauge field to the inflaton.  For the sake of generality, we will not specify explicit forms of the inflaton potential and the coupling functions $I_{1,2}$ (we shall do that
in Sec.~\ref{sec-nonminimal}). The kinetic coupling function $I_1(\phi)$ must be (i)~positive in order to ensure the positive-definiteness of the gauge-field energy density and (ii)~always greater than unity during inflation in order to avoid the strong coupling problem. On the other hand, the axial coupling function $I_2(\phi)$ may be completely arbitrary because it does not enter the energy-momentum tensor [see Eq.~(\ref{EMT-mixed}) below] and does not have any impact of the coupling of other matter to the gauge field.
		
Varying action~(\ref{action-kinetic-axial}) with respect to the gauge field, we get the Maxwell equations of the following form:
\begin{equation}
\label{Maxwell}
\frac{1}{\sqrt{-g}}\partial_{\mu}\left[\sqrt{-g}I_1(\phi)F^{\mu\nu}\right]=-\tilde{F}^{\mu\nu}\partial_\mu I_2(\phi)
\end{equation}
They are supplemented by the Bianchi identities for the dual tensor
\begin{equation}
\frac{1}{\sqrt{-g}}\partial_{\mu}\left[\sqrt{-g}\tilde{F}^{\mu\nu}\right]=0.
\end{equation}

The equation of motion for the inflaton field can be found by varying the action in Eq.~(\ref{action-kinetic-axial}) with respect to $\phi$:
\begin{multline}
\label{KGF}
\frac{1}{\sqrt{-g}}\partial_{\mu}\left[\sqrt{-g}g^{\mu\nu}\partial_{\nu}\phi\right]+\frac{dV}{d\phi}=\\
-\frac{1}{4}\frac{dI_1}{d\phi}\langle F_{\mu\nu}F^{\mu\nu}\rangle-\frac{1}{4}\frac{dI_2}{d\phi}\langle F_{\mu\nu}\tilde{F}^{\mu\nu}\rangle,
\end{multline}
where we take into account that the inflaton is a classical field while the gauge field is considered as a quantum field; therefore, the vacuum expectation value $\langle...\rangle$ is taken on the right-hand side.

We now assume that during inflation spacetime is a spatially-flat Friedmann-Lema\^{i}tre-Robertson-Walker (FLRW) universe. In terms of cosmic time $t$ it reads as
\begin{equation}
\label{metric}
g_{\mu\nu}=\mathrm{diag}\,(1,\,-a^{2},\,-a^{2},\,-a^{2}),
\end{equation}
and thus $\sqrt{-g}=a^{3}$. Here $a(t)$ is the scale factor which describes the expansion of the Universe. Moreover, we assume that the inflaton field is spatially homogeneous and only depends on time $\phi=\phi(t)$. As for the gauge field, we introduce the three-vectors of the electric $\boldsymbol{E}$ and the magnetic $\boldsymbol{B}$ fields~\footnote{Throughout this work we use the terms ``electric'' and ``magnetic'' to denote the gauge-field quantities which correspond to the usual electric and magnetic fields in electrodynamics. This, however, does not restrict our analysis to the case of electromagnetic field; $A_{\mu}$ is an arbitrary Abelian gauge field.} which parametrize the gauge-field tensor and its dual as follows:
\begin{equation}
\label{EM-tensors}
\arraycolsep=7pt
\begin{array}{ll}
 F^{0i}=\dfrac{1}{a} E^{i},& F_{ij}=a^{2}\varepsilon_{ijk}B^{k},\\[10pt] \vspace*{3mm}
\tilde{F}^{0i}=\dfrac{1}{a}B^{i}, & \tilde{F}_{ij}=-a^{2}\varepsilon_{ijk}E^{k},
\end{array}
\end{equation}
where $\varepsilon_{ijk}$ is the three-dimensional Levi-Civita symbol and indices $i,\,j,\,k$ on the right-hand side denote the components of
three-vectors; i.e., they are raised and lowered by the Euclidean metric. The powers of the scale factor $a(t)$ are chosen  such that the quantities $\boldsymbol{E}$ and $\boldsymbol{B}$ represent the physical fields measured by a comoving observer.

Then, the Maxwell equations expressed in terms of the electric and magnetic fields take the form:
\begin{eqnarray}
	&&\dot{\boldsymbol{E}}+2H\boldsymbol{E}-\frac{1}{a}\mathrm{rot\,}\boldsymbol{B}=-\frac{\dot{I}_{1}}{I_{1}}\boldsymbol{E}-\frac{\dot{I}_{2}}{I_{1}}\boldsymbol{B},\label{Maxwell-2}\\
	&&\dot{\boldsymbol{B}}+2H\boldsymbol{B}+\frac{1}{a}\mathrm{rot\,}\boldsymbol{E}=0,\label{Maxwell-second-equation}\\
	&&\mathrm{div\,}\boldsymbol{E}=0,\qquad \mathrm{div\,}\boldsymbol{B}=0,
\end{eqnarray}
where the overdot denotes a derivative with respect to the
cosmic time $t$ and $H=\frac{\dot{a}}{a}$ is the Hubble parameter. 

The Klein-Gordon equation for the inflaton field becomes
\begin{equation}
\label{KGF-2}
\ddot{\phi}+3H\dot{\phi}+\frac{dV}{d\phi}=\frac{1}{2}\frac{dI_1}{d\phi}\langle \boldsymbol{E}^{2}-\boldsymbol{B}^{2}\rangle+\frac{dI_2}{d\phi}\langle \boldsymbol{E}\cdot\boldsymbol{B}\rangle.
\end{equation}

It is important to note that Eqs.~(\ref{Maxwell-2})--(\ref{KGF-2}) form a coupled system of differential equations. On one hand, the inflaton field enters the Maxwell equation through the kinetic and axial coupling functions $I_{1,2}$ triggering the gauge-field production (remember that in the absence of kinetic and axial couplings the conformal invariance prevents the gauge field from amplification~\cite{Parker:1968}). On the other hand, the gauge field appears in the Klein-Gordon equation causing backreaction on the slow-roll evolution of the inflaton. If the generated gauge field is strong enough to make the terms on the right-hand side of Eq.~(\ref{KGF-2}) comparable to those on the left-hand side, then both inflaton and gauge fields must be treated self-consistently, i.e., on the same footing.

According to the Friedmann equation, the expansion rate 
 of the Universe is determined by the total energy density $\rho$ of all matter fields:
\begin{eqnarray}
\label{Friedmann}
H^{2}=\left(\frac{\dot{a}}{a}\right)^{2}=\frac{1}{3M_{\mathrm{P}}^{2}}\rho,
\end{eqnarray} 
The latter can be calculated as the $00$-component of the energy-momentum tensor which is found by varying action~(\ref{action-kinetic-axial}) with respect to metric:
\begin{multline}
\label{EMT-mixed}
T_{\mu\nu}=\frac{2}{\sqrt{-g}}\frac{\delta S_{KA}}{\delta g^{\mu\nu}}=\partial_{\mu}\phi\partial_{\nu}\phi-
I_1(\phi)g^{\alpha\beta}\langle F_{\mu\alpha}F_{\nu\beta}\rangle\\
-g_{\mu\nu} \left[\frac{1}{2}g^{\alpha\beta}\partial_{\alpha}\phi\partial_{\beta}\phi-V(\phi)-\frac{1}{4}I_1(\phi)\langle F_{\alpha\beta}F^{\alpha\beta}\rangle\right].
\end{multline}
Here again, we take the vacuum expectation value of the gauge-field contribution since it is considered as a quantum field while the energy-momentum tensor is a classical object. Then, the energy density reads 
\begin{equation}
\label{energy-density}
\rho=T_{00}=\left[\frac{1}{2}\dot{\phi}^{2}+V(\phi)\right]+\frac{1}{2}I_1(\phi)\langle \boldsymbol{E}^{2}+\boldsymbol{B}^{2}\rangle =\rho_\mathrm{inf}+\rho_\mathrm{GF}.
\end{equation}
Note that only the kinetic coupling $I_1$ appears in the energy density. Another important comment is that the backreaction of the produced gauge field arises also in the Friedmann  equation if $\rho_\mathrm{GF}$ becomes comparable to $\rho_\mathrm{inf}$. This may lead to a big deviation of the equation of state from the vacuumlike behavior $w\approx -1$ and, thus, terminate inflation. 

Therefore, it is important to keep track of backreaction of the generated fields on the background evolution since it may significantly modify the dynamics of the expansion and change the predictions for observables such as, e.g., the power spectra of primordial perturbations. In order to do this, one needs a tool which can take into account these nonlinear effects fully self-consistently. One such tool\,---\,the gradient expansion formalism\,---\,already exists in the cases of purely kinetic \cite{Sobol:2018,Sobol:2020lec}, or purely axial \cite{Sobol:2019,Gorbar:2021rlt} couplings. In the next section we generalize this approach to the case when both types of coupling are present.

\section{Gradient expansion formalism}
\label{sec-gef}

\subsection{System of equations for bilinear gauge-field correlators}
\label{sec-blf}

The gauge field, as any other matter field, exists during inflation in the form of vacuum fluctuations. The kinetic and axial couplings to the inflaton field break the conformal invariance of the gauge-field action and thus enable the amplification of its Fourier modes when they cross the horizon, in a similar manner to the generation of primordial perturbations. Modes with  wavelengths largely exceeding the radius of observable region behave as classical mean fields; however, their quantum origin implies that they are stochastic quantities; i.e., they are chaotically oriented in different regions of the Universe. This means that vector quantities as $\boldsymbol{E}$ or $\boldsymbol{B}$ average to zero and are not suitable for the description of the generated fields. It is much more convenient to use a set of scalar quantities which are the vacuum expectation values of different scalar products of $\boldsymbol{E}$, $\boldsymbol{B}$ and their spatial derivatives (curls). Thus, we introduce the following bilinear gauge-field quantities:
\begin{eqnarray}
\label{EE}
\mathscr{E}^{(n)}&\equiv&\frac{I_1(\phi)}{a^{n}}\langle \boldsymbol{E}\cdot \mathrm{rot}^{n}\boldsymbol{E}\rangle,\\
\label{EB}
\mathscr{G}^{(n)}&\equiv&-\frac{I_1(\phi)}{2a^{n}}\langle \boldsymbol{E}\cdot \mathrm{rot}^{n}\boldsymbol{B}
+\mathrm{rot}^{n}\boldsymbol{B}\cdot \boldsymbol{E}
\rangle,\\
\label{BB}
\mathscr{B}^{(n)}&\equiv&\frac{I_1(\phi)}{a^{n}}\langle \boldsymbol{B}\cdot \mathrm{rot}^{n}\boldsymbol{B}\rangle,
\end{eqnarray}
where $\langle \ldots \rangle$ denote the vacuum expectation value. 
Using Maxwell's equations~(\ref{Maxwell-2}), (\ref{Maxwell-second-equation}), we derive a system of coupled equations for these quantities:
\begin{multline}
\dot{\mathscr{E}}^{(n)}+(n+4)H\mathscr{E}^{(n)}+\frac{\dot{I_1}}{I_1}\mathscr{E}^{(n)}-2\frac{\dot{I_2}}{I_1}\mathscr{G}^{(n)}\\
+2\mathscr{G}^{(n+1)}=[\dot{\mathscr{E}}^{(n)}]_{\mathrm{b}},\label{eq-EE}
\end{multline}
\begin{multline}
 \dot{\mathscr{G}}^{(n)}+(n+4)H\mathscr{G}^{(n)}
-\frac{\dot{I_2}}{I_1}\mathscr{B}^{(n)}
+\mathscr{B}^{(n+1)}\\
-\mathscr{E}^{(n+1)}=[\dot{\mathscr{G}}^{(n)}]_{\mathrm{b}}. \label{eq-EB}
\end{multline}
\begin{multline}
\dot{\mathscr{B}}^{(n)} + (n+4)H\,	\mathscr{B}^{(n)}-\frac{\dot{I_1}}{I_1}\mathscr{B}^{(n)}-2\mathscr{G}^{(n+1)}\\
=[\dot{\mathscr{B}}^{(n)}]_{\mathrm{b}}. \label{eq-BB}
\end{multline}
First, we note that direct application of Maxwell's equations (\ref{Maxwell-2})--(\ref{Maxwell-second-equation}) to the gauge-field correlators (\ref{EE})--(\ref{BB}) gives   equations \eqref{eq-EE} to \eqref{eq-BB} above with vanishing right-hand sides. This would  indeed be correct if the functions~(\ref{EE})--(\ref{BB}) would include the contributions of all Fourier modes of the gauge field. However, this would mix the finite and physically meaningful contributions of the amplified gauge-field modes due to kinetic and axial coupling and the (infinite) contributions of vacuum fluctuations which are present also in the absence of any coupling to the inflaton. In order to exclude the latter contribution (a simple way or ``renormalization''), in the next subsection we will specify the range of Fourier modes which are physically relevant and we show that the number of these modes typically grows during inflation. Therefore, we must take into account the change of bilinear functions also due to the fact that the number of modes is changing. This is precisely the reason of introducing the extra terms on the right-hand sides of Eqs.~(\ref{eq-EE})--(\ref{eq-BB}), the boundary terms.

Another important comment is that any equation of motion for the $n$th-order functions always contains at least one function with the $(n+1)$th power of the curl. As a result, all equations in the system~(\ref{eq-EE})--(\ref{eq-BB}) are coupled into an infinite hierarchy. For practical purposes, one must truncate this chain at some finite order. In the end of this section we suggest a possible way for this truncation.

\subsection{Origin of boundary terms}
\label{sec-bd1}

Our goal is to describe the gauge field generation during inflation by kinetic and axial coupling to the inflaton field. First, it is important to choose the observables which quantify this process. As we discussed in the previous subsection, these can be the bilinear gauge-field correlators defined in Eqs.~(\ref{EE})--(\ref{BB}). It is convenient to compute the vacuum expectation values using the decomposition of gauge-field operators over the set of annihilation and creation operators in Fourier space. In Coulomb gauge, $A_{\mu}=(0,\,\boldsymbol{A})$ and $\mathrm{div\,}\boldsymbol{A}=0$, the gauge field operator has the form
\begin{multline}
\label{quantized_A}
\bm{A}(t,\bm{x})=\int\frac{d^{3}\bm{k}}{(2\pi)^{3/2}\sqrt{I_1}}\sum_{\lambda=\pm}\Big[\boldsymbol{\epsilon}^{\lambda}(\bm{k})\hat{a}_{\bm{k},\lambda}A_{\lambda}(t,k)e^{i\bm{k}\cdot\bm{x}}\\
+\boldsymbol{\epsilon}^{*\lambda}(\bm{k})\hat{a}_{\bm{k},\lambda}^{\dagger}A_{\lambda}^{*}(t,k)e^{-i\bm{k}\cdot\bm{x}} \Big],
\end{multline}
where $A_{\lambda}(t,k)$ is the mode function, $\boldsymbol{\epsilon}^{\lambda}(\bm{k})$ is the polarization three-vector, $\hat{a}_{\bm{k},\lambda}$ ($\hat{a}^{\dagger}_{\bm{k},\lambda}$) is the annihilation (creation) operator of the electromagnetic mode with momentum $\bm{k}$ and circular polarization $\lambda=\pm$, and $k=|\bm{k}|$. Note that for further convenience we included a factor $1/\sqrt{I_1}$ in decomposition~(\ref{quantized_A}). Polarization vectors are transverse $\bm{k}\cdot\boldsymbol{\epsilon}^{\lambda}(\bm{k})=0$ and have the following properties:
\begin{equation}
\begin{array}{l}
   \boldsymbol{\epsilon}^{*\lambda}(\bm{k})=\boldsymbol{\epsilon}^{-\lambda}(\bm{k}), \quad [i\bm{k}\times\boldsymbol{\epsilon}^{\lambda}(\bm{k})]=\lambda k \boldsymbol{\epsilon}^{\lambda}(\bm{k}), \\[10pt] \boldsymbol{\epsilon}^{*\lambda}(\bm{k})\cdot\boldsymbol{\epsilon}^{\lambda'}(\bm{k})=\delta^{\lambda\lambda'}.
\end{array}
\end{equation}
The creation and annihilation operators satisfy the canonical commutation relations
\begin{equation}
[\hat{a}_{\bm{k},\lambda},\,\hat{a}^{\dagger}_{\bm{k}',\lambda'}]=\delta_{\lambda\lambda'}\delta^{(3)}(\bm{k}-\bm{k}').
\end{equation}

The electric and magnetic field operators are given by
\begin{equation}
\label{fields-E-and-B}
\boldsymbol{E}= -\frac{1}{a}\dot{\boldsymbol{A}}, \qquad \boldsymbol{B}= \frac{1}{a^{2}} \mathrm{rot\,}\boldsymbol{A}.
\end{equation}
Using decomposition (\ref{quantized_A}) it is straightforward to find the corresponding results for $\boldsymbol{E}$ and $\boldsymbol{B}$. Substituting them into expressions for bilinear functions (\ref{EE})--(\ref{BB}), we obtain the following spectral representations:
\begin{equation}
\label{E_1}
\mathscr{E}^{(n)}=\sum_{\lambda=\pm 1}\int_{0}^{k_{\mathrm{h}}}\frac{d k}{k} \lambda^{n}\frac{k^{n+3}I_1}{2\pi^{2}a^{n+2}}\,\Big|\frac{d}{dt}\Big(\frac{A_{\lambda}(t,k)}{\sqrt{I_{1}}}\Big)\Big|^{2},
\end{equation}
\begin{equation}
\label{G_1}
\mathscr{G}^{(n)}=\sum_{\lambda=\pm 1}^{}\int_{0}^{k_{\mathrm{h}}}\frac{d k}{k} \lambda^{n+1}\frac{k^{n+4}I_1}{4\pi^{2}a^{n+3}}\,\frac{d}{d t}\Big|\frac{A_{\lambda}(t,k)}{\sqrt{I_{1}}}\Big|^{2},
\end{equation}
\begin{equation}
\label{B_1}
\mathscr{B}^{(n)}=\sum_{\lambda=\pm 1}^{}\int_{0}^{k_{\mathrm{h}}}\frac{d k}{k} \lambda^{n}\frac{k^{n+5}}{2\pi^{2}a^{n+4}}\,|A_{\lambda}(t,k)|^{2}.
\end{equation}

In general, the vacuum expectation value implies that all Fourier modes (with wave numbers from zero to infinity) must be taken into account in decompositions~(\ref{E_1})--(\ref{B_1}). However, as we will see below, not all gauge-field modes are enhanced during inflation and, thus, not all of them should be included if we want to consider only the gauge field generated due to kinetic and axial couplings. On  general grounds, it is clear that the modes with arbitrarily large momentum, $k\to\infty$ should not be affected by the coupling to the inflaton field, because the energy required to generate them is unbounded from above. Therefore, there always exists some threshold momentum\footnote{By analogy with the theory of primordial perturbations, we will associate this momentum with horizon crossing (explains the index ``$\mathrm{h}$'' in $k_{\mathrm{h}}$). However, now this is not the standard Hubble horizon, but some effective horizon for the gauge field. We will see below that this horizon separates the modes which undergo a tachyonic instability from the ones which always oscillate in time.}, $k_{\mathrm{h}}$, above which the modes can be considered as pure vacuum fluctuations. Naturally, we should take into account only the modes with momenta $k<k_{\mathrm{h}}$. This explains the upper boundary in the integrals~(\ref{E_1})--(\ref{B_1}).

If this threshold momentum depends on time, $k_{\mathrm{h}}=k_{\mathrm{h}}(t)$, the quantities $\mathscr{E}^{(n)}$, $\mathscr{G}^{(n)}$, $\mathscr{B}^{(n)}$ obtain additional time dependence: they change not only because the corresponding spectral densities are time-dependent, but also due to the time-varying upper integration boundary. The latter time dependence is not captured by Maxwell's equations since this mode separation was introduced manually. In order to compensate this additional time dependence, we have to introduce extra terms in the corresponding equations of motion which are the boundary terms.

If a general bilinear function $X\in\{\mathscr{E}^{(n)},\ \mathscr{G}^{(n)},\ \mathscr{B}^{(n)}\}$ is represented by its spectral decomposition
\begin{equation}
X=\sum_{\lambda=\pm 1}^{}\int_{0}^{k_{\mathrm{h}}(t)}\frac{d k}{k}\frac{dX(\lambda,k)}{d\ln k},
\end{equation}
the boundary term in its equation of motion has the form:
\begin{equation}
\label{bt-general}
[\dot{X}]_{\mathrm{b}}=\sum_{\lambda=\pm 1}^{}\frac{dX(\lambda,k)}{d\ln k}\Big|_{k=k_{\mathrm{h}}} \times \frac{d\ln k_{\mathrm{h}}}{dt}.
\end{equation}
Thus, in order to compute the boundary terms in Eqs.~(\ref{eq-EE})--(\ref{eq-BB}), we need to know the spectral densities of all quantities $\mathscr{E}^{(n)}$, $\mathscr{G}^{(n)}$, $\mathscr{B}^{(n)}$ at the threshold momentum and the time dependence of the threshold momentum itself. All this will be determined in the next subsection.

\subsection{Time evolution of mode functions}
\label{sec-mode}

Now let us study the time evolution of the gauge-field modes during inflation. Substituting Eq.~(\ref{quantized_A}) into Eq.~(\ref{Maxwell-2}), we get the following equation for the mode function $A_{\lambda}(t,k)$:
\begin{multline}
\label{eq-mode-2}
\ddot{A}_{\lambda}(t,k)+H\dot{A}_{\lambda}(t,k)+A_{\lambda}(t,k)\times\\
\times
\left(\frac{k^{2}}{a^{2}}-H\frac{\dot{I}_{1}}{2I_{1}}-
\frac{\ddot{I_1}}{2I_1}+\frac{\dot{I}_{1}^2}{4I_{1}^2}
-\lambda\frac{k}{a}\frac{\dot{I}_{2}}{I_{1}}\right)=0.
\end{multline}
We now switch to the new variable $z=k\eta$, where $\eta=\int^{t}dt'/a(t')$ is  conformal time. Also, for the sake of brevity, we introduce the following functions:
\begin{equation}
\label{xi-s}
\xi(t)=\frac{\dot{I}_{2}}{2H\,I_{1}}, \qquad 
s(t)=\frac{\dot{I}_{1}}{2H I_{1}}+
\frac{\ddot{I_1}}{2H^2 I_1}-\frac{\dot{I}_{1}^2}{4H^2 I_{1}^2},
\end{equation}
which both  are functions of time (or, equivalently, $z$) during inflation. The mode equation~(\ref{eq-mode-2}) then becomes
\begin{equation}
\label{eq-mode-z-1}
\frac{d^2 A_{\lambda}(z,k)}{dz^2}+\Big[1-2\lambda \frac{aH}{k} \xi(z)-\Big(\frac{aH}{k}\Big)^2 s(z) \Big]A_{\lambda}(z,k)=0.
\end{equation}
Finally, we note that during inflation the conformal time can be approximately expressed as
\begin{equation}
\label{eta-de-Sitter}
\eta\simeq -1/(aH),
\end{equation}
which corresponds to the pure de Sitter solution. Although this is not an exact relation in realistic inflationary models, the deviations from it are suppressed by the slow-roll parameters and for a relatively short time interval Eq.~(\ref{eta-de-Sitter}) may be used. We this, we finally arrive at the mode equation 
\begin{equation}
\label{eq-mode-z-2}
\frac{d^2 A_{\lambda}(z,k)}{dz^2}+\Big[1+\lambda \frac{2\xi(z)}{z}-\frac{s(z)}{z^2} \Big]A_{\lambda}(z,k)=0.
\end{equation}

During inflation, the conformal time (and variable $z$) grows from large negative values to zero; i.e., its absolute value always decreases. Therefore, at very early times, i.e., for sufficiently large $|z|\gg |2\xi|$ and $|z|\gg\sqrt{|s|}$, first term in the bracket\,---\,the unity\,---\,dominates over the other two. In this case, we obtain the equation of motion of a harmonic oscillator with unit frequency. Thus, the mode function oscillates in conformal time in the same way as it would behave in Minkowski spacetime and corresponds to vacuum fluctuations of the gauge field. Therefore, it is possible to impose the boundary condition for the mode function in the form of Bunch-Davies vacuum solution \cite{Bunch:1978}:
\begin{equation}
\label{BD}
A_{\lambda}(z,k)=\frac{1}{\sqrt{2k}}e^{-iz}, \quad -z\gg 1.
\end{equation}
In this regime (which we call ``subhorizon'') the effect of kinetic and axial couplings is negligible because they appear only in the second and third terms in the bracket of Eq.~(\ref{eq-mode-z-2}) which are suppressed at large $z$. Therefore, subhorizon modes are not relevant for the description of magnetogenesis.

In the opposite case, $|z|\ll |2\xi|$ and/or $|z|\ll \sqrt{|s|}$, the first term is negligible and the time evolution of the mode function is fully determined by its coupling to the inflaton field. Depending on the sign of $s$, $\xi$, and $\lambda$, the expression in brackets in Eq.~(\ref{eq-mode-z-2}) may become negative and a tachyonic instability may occur for the corresponding Fourier mode. In this case, the mode function is exponentially amplified. However, even if the mode is not tachyonically unstable, its time evolution in this regime (called superhorizon) strongly differs from that of vacuum fluctuations. Therefore, such superhorizon modes must be taken into account in the generated gauge field.

An important practical question is how to determine the threshold momentum $k_{\mathrm{h}}$ which separates sub- and superhorizon modes. In the literature it is usually chosen as  the condition that the first term in brackets in Eq.~(\ref{eq-mode-z-2}) becomes comparable to the absolute value of at least one of the other terms. Namely, if the absolute value of the second term in brackets is greater then the one of the third term (as, e.g., in the case of pure axial coupling), the natural choice for the threshold momentum is
\begin{equation}
\label{k-h-axial}
    k_{\mathrm{h}}=2|\xi|aH,
\end{equation}
while in the opposite situation when the third term is dominant,
\begin{equation}
\label{k-h-kinetic}
    k_{\mathrm{h}}=\sqrt{|s|}aH.
\end{equation}
In the most general case, when both terms are important, one can take a smooth interpolation between expressions in Eqs.~(\ref{k-h-axial}) and (\ref{k-h-kinetic}). One possible option is to take the greater solution of quadratic equation
\begin{equation}
    \Big(\frac{k}{aH}\Big)^2-2|\xi|\frac{k}{aH} -|s|=0
\end{equation}
that is
\begin{equation}
\label{k-h-mixed}
    k_{\mathrm{h}}=aH\big(|\xi|+\sqrt{\xi^2 + |s|}\big).
\end{equation}
Finally, if the expression for $k_{\mathrm{h}}$ in Eq.~(\ref{k-h-mixed}) is not growing monotonically during inflation, we can take the upper monotonic envelope of this function, i.e., at any moment of time $t$ we use its maximal value over all preceding moments of time. The final expression for the threshold momentum that will be used in the rest of the article, then is of the form
\begin{equation}
\label{k-h-fin}
    k_{\mathrm{h}}(t)=\max\limits_{t'\leq t}\Big\{a(t')H(t')\Big[|\xi(t')|+\sqrt{\xi^2(t')+|s(t')|} \Big] \Big\}.
\end{equation}
If $k=k_{\mathrm{h}}(t)$, we say that the gauge-field mode with momentum $k$ crosses the horizon at the moment of time $t$.

It is easy to see that during inflation $k_{\mathrm{h}}(t)$ typically grows. Indeed, the Hubble parameter $H(t)$ is a slowly varying function ($\epsilon_H=-\dot{H}/H^2 \ll 1$ during inflation). Although it is impossible to predict the behavior of functions $\xi(t)$ and $s(t)$ for arbitrary coupling functions $I_{1,2}$, we note that they are proportional to $\dot{\phi}$ and $\ddot{\phi}$; therefore, their time evolution is not expected to be fast. Taking into account that the scale factor $a(t)$ grows nearly exponentially during inflation, we conclude that the threshold momentum $k_{\mathrm{h}}(t)$ rapidly increases in time (maybe except for some short intervals of time when functions $\xi$ and $s$ quickly decrease or cross zero). 

Thus, we see that the number of physically relevant gauge-field modes grows during inflation which is precisely the reason for appearance of boundary terms in Eqs.~(\ref{eq-EE})--(\ref{eq-BB}). In order to find their explicit form, we need to know the mode function in a small vicinity of time around the moment of horizon crossing. For a given momentum $k$, this moment of time $t_{\mathrm{h}}(k)$ can be found by inverting Eq.~(\ref{k-h-fin}). For a short time interval around $t_{\mathrm{h}}$, we may assume that functions $\xi$ and $s$ in Eq.~(\ref{eq-mode-z-1}) are constant: $\xi\approx\xi(t_{\mathrm{h}})= \text{const}$, $s\approx s(t_{\mathrm{h}})=\text{const}$. In this case, Eq.~(\ref{eq-mode-z-2}) admits a solution in terms of Whittaker functions 
$W_{\varkappa,\mu}(2iz)$ (for details, see Appendix~\ref{app-Whittaker}). Taking into account the Bunch-Davies boundary condition~(\ref{BD}), we obtain the positive-frequency solution of the mode equation in the following form:
\begin{equation}
\label{A_Whittaker}
A_{\lambda}(z,k)=\frac{1}{\sqrt{2k}}e^{\frac{\pi\lambda \xi(t_{\mathrm{h}})}{2}}
\,W_{\varkappa,\mu}(2iz),
\end{equation}
$$
\varkappa=-i\lambda \xi(t_{\mathrm{h}}), \quad \mu=\sqrt{\frac{1}{4}+s(t_{\mathrm{h}})}.
$$
The derivative of this expression with respect to $z$ can be computed by using Eqs.~(\ref{Whittaker-eq})--(\ref{W-asym}) in Appendix~\ref{app-Whittaker}.

\subsection{Exact expressions for the boundary terms}
\label{sec-bd}

Having determined the approximate expression for the mode function close to the moment of horizon crossing in the previous subsection, we are now ready to compute the boundary terms in Eqs.~(\ref{eq-EE})--(\ref{eq-BB}). Using the general form of the boundary term in Eq.~(\ref{bt-general}) and spectral densities of $\mathscr{E}^{(n)}$, $\mathscr{G}^{(n)}$, $\mathscr{B}^{(n)}$ in Eqs.~(\ref{E_1})--(\ref{B_1}), we obtain the following expressions:
\begin{multline}
\label{E_p_d}
[\dot{\mathscr{E}}^{(n)}]_{\mathrm{b}}=\frac{d \ln k_{\mathrm{h}}(t)}{d t}\frac{1}{4\pi^{2}}\left(\frac{k_{\mathrm{h}}(t)}{a(t)}\right)^{n+4}\\
\times \sum_{\lambda=\pm 1}\lambda^{n} E_{\lambda}(\xi(t),\,s(t),\,\sigma(t)),
\end{multline}
\begin{multline}
\label{G_p_d}
[\dot{\mathscr{G}}^{(n)}]_{\mathrm{b}}=\frac{d \ln k_{\mathrm{h}}(t)}{d t}\frac{1}{4\pi^{2}}\left(\frac{k_{\mathrm{h}}(t)}{a(t)}\right)^{n+4}\\
\times\sum_{\lambda=\pm 1}\lambda^{n+1} G_{\lambda}(\xi(t),\,s(t),\,\sigma(t)),
\end{multline}
\begin{multline}
\label{B_p_d}
[\dot{\mathscr{B}}^{(n)}]_{\mathrm{b}}=\frac{d \ln k_{\mathrm{h}}(t)}{d t}\frac{1}{4\pi^{2}}\left(\frac{k_{\mathrm{h}}(t)}{a(t)}\right)^{n+4}\\
\times\sum_{\lambda=\pm 1}\lambda^{n} B_{\lambda}(\xi(t),\,s(t)),
\end{multline}
where the threshold momentum $k_{\mathrm{h}}(t)$ is given by Eq.~(\ref{k-h-fin}) and
\begin{multline}
\label{E-lambda}
E_{\lambda}(\xi,\,s,\,\sigma)=\frac{e^{\pi\lambda \xi}}{r^2} \Big|\left(i r-i\lambda \xi-\sigma\right)W_{-i\lambda\xi,\sqrt{s+\frac{1}{4}}}(-2i r)\\
+W_{1-i\lambda\xi,\sqrt{s+\frac{1}{4}}}(-2i r)\Big|^{2},
\end{multline}
\begin{multline}
\label{G-lambda}
G_{\lambda}(\xi,\,s,\,\sigma)=\frac{e^{\pi\lambda \xi}}{r}\Big\{-\sigma \Big|W_{-i\lambda\xi,\sqrt{s+\frac{1}{4}}}(-2i r) \Big|^{2}\\
+\Re e\Big[W_{i\lambda \xi,\sqrt{s+\frac{1}{4}}}(2i r) W_{1-i\lambda\xi,\sqrt{s+\frac{1}{4}}}(-2i r)\Big]\Big\},
\end{multline}
\begin{equation}
\label{B-lambda}
    \qquad B_{\lambda}(\xi,\,s)=e^{\pi\lambda \xi}\,\left|W_{-i\lambda\xi,\sqrt{s+\frac{1}{4}}}(-2i r) \right|^{2}.
\end{equation}
Here $r=r(\xi,\,s)=|\xi|+\sqrt{\xi^{2}+|s|}$ and a new function $\sigma(t)$ has been introduced:
\begin{equation}
\label{sigma-par}
    \sigma(t)=\frac{\dot{I}_1}{2H I_1}.
\end{equation}
Note that everywhere in the functions $E_\lambda$, $G_\lambda$, and $B_\lambda$ we use Eq.~(\ref{k-h-mixed}) for the horizon-crossing time instead of Eq.~(\ref{k-h-fin}); i.e., we set $z_{\mathrm{h}}=k_{\mathrm{h}} \eta\approx -k_{\mathrm{h}}/(aH)=-r(\xi,\,s)$. This can be justified by the following observation. If $k_{\mathrm{h}}(t)$ is growing in time ($d\ln k_{\mathrm{h}}/dt>0$) both expressions, (\ref{k-h-mixed}) and (\ref{k-h-fin}) coincide. On the other hand, when $k_{\mathrm{h}}(t)$ is constant, Eq.~(\ref{k-h-mixed}) gives a smaller value than Eq.~(\ref{k-h-fin}); however, total expression for the boundary term vanishes because $d\ln k_{\mathrm{h}}/dt=0$ and it makes no difference which expression for $k_{\mathrm{h}}$ is used inside the functions $E_\lambda$, $G_\lambda$, and $B_\lambda$.

\subsection{Full system of equations}

Finally, we find the following closed system of equations, which determines the time evolution 
of the bilinear gauge-field functions, scale factor and the inflation field:
\begin{equation}
H^{2}=\frac{1}{3M_{\mathrm{P}}^{2}}\left[\frac{1}{2}\dot{\phi}^{2}+V(\phi)+\frac{1}{2}\left(\mathscr{E}^{(0)}+\mathscr{B}^{(0)}\right)\right],
\label{Friedmann-fin} 
\end{equation}
\begin{equation}
\label{KGF-3}
\ddot{\phi}+3H\dot{\phi}+\frac{dV}{d\phi}=
\frac{1}{2I_1}\frac{dI_1}{d\phi}\left(\mathscr{E}^{(0)}-\mathscr{B}^{(0)}\right)-
\frac{1}{I_1}\frac{dI_2}{d\phi}\mathscr{G}^{(0)}.
\end{equation}
\begin{multline}
\dot{\mathscr{E}}^{(n)}+(n+4)H\mathscr{E}^{(n)}+\frac{\dot{I_1}}{I_1}\mathscr{E}^{(n)}-2\frac{\dot{I_2}}{I_1}\mathscr{G}^{(n)}+2\mathscr{G}^{(n+1)}\\
=\frac{d \ln k_{\mathrm{h}}}{d t}\frac{1}{4\pi^{2}}\left(\frac{k_{\mathrm{h}}}{a}\right)^{n+4}\sum_{\lambda=\pm 1}\lambda^{n} E_{\lambda}(\xi,\,s,\,\sigma),
\label{eq-EE1}
\end{multline}
\begin{multline}
\dot{\mathscr{G}}^{(n)}+(n+4)H\mathscr{G}^{(n)}
-\frac{\dot{I_2}}{I_1}\mathscr{B}^{(n)}
+\mathscr{B}^{(n+1)}-\mathscr{E}^{(n+1)}\\
=\frac{d \ln k_{\mathrm{h}}}{d t}\frac{1}{4\pi^{2}}\left(\frac{k_{\mathrm{h}}}{a}\right)^{n+4}\sum_{\lambda=\pm 1}\lambda^{n+1}G_{\lambda}(\xi,\,s,\sigma), 
\label{eq-EB1}
\end{multline}
\begin{multline}
\dot{\mathscr{B}}^{(n)} + (n+4)H\,	\mathscr{B}^{(n)}-\frac{\dot{I_1}}{I_1}\mathscr{B}^{(n)}-2\mathscr{G}^{(n+1)}\\
=\frac{d \ln k_{\mathrm{h}}}{d t}\frac{1}{4\pi^{2}}\left(\frac{k_{\mathrm{h}}}{a}\right)^{n+4}\sum_{\lambda=\pm 1}\lambda^{n}B_{\lambda}(\xi,\,s),\label{eq-BB1}
\end{multline}
where $k_{\mathrm{h}}(t)$ is given by Eq.~(\ref{k-h-fin}) 
and $E_{\lambda}(\xi,\,s,\,\sigma)$, $G_{\lambda}(\xi,\,s,\,\sigma)$, and $B_{\lambda}(\xi,\,s)$ are given in Eqs.~(\ref{E-lambda})--(\ref{B-lambda}).

In order to get more insight in the physical processes at work in this system of equations, let us consider the time evolution of the inflaton energy density $\rho_{\mathrm{inf}}=\dot{\phi}^2/2+V(\phi)$ and the gauge-field energy density $\rho_{\mathrm{GF}}=(\mathscr{E}^{(0)}+\mathscr{B}^{(0)})/2$. For the former quantity, we multiply the Klein-Gordon equation~(\ref{KGF-3}) by $\dot{\phi}$ and rewrite it in the following form:
\begin{equation}
\label{eq-evol-rho-inf}
\dot{\rho}_{\mathrm{inf}}+3H(\rho_{\mathrm{inf}}+p_{\mathrm{inf}})=-\frac{\dot{I}_1}{2I_1}(\mathscr{B}^{(0)}-\mathscr{E}^{(0)})-\frac{\dot{I}_2}{I_1}\mathscr{G}^{(0)},
\end{equation}
where $p_{\mathrm{inf}}=\dot{\phi}^2/2-V(\phi)$ is the pressure of the inflaton field. Note that this equation has the form of a covariant energy conservation for the inflaton field with a source on the right-hand side. The physical meaning of this source term becomes clear when we write down the corresponding equation for the gauge-field energy density. Taking half of the sum of Eqs.~(\ref{eq-EE1}) and (\ref{eq-BB1}) for $n=0$, we get
\begin{equation}
\label{eq-evol-rho-GF}
\dot{\rho}_{\mathrm{GF}}+3H(\rho_{\mathrm{GF}}+p_{\mathrm{GF}})= \frac{\dot{I}_1}{2I_1}(\mathscr{B}^{(0)}-\mathscr{E}^{(0)})+\frac{\dot{I}_2}{I_1}\mathscr{G}^{(0)}+[\dot{\rho}_{\mathrm{GF}}]_\mathrm{b}.
\end{equation}
Here $p_{\mathrm{GF}}=(1/3)\rho_{\mathrm{GF}}$ is the gauge field pressure as for the radiation. Again, this has the form of an equation of covariant energy conservation where the first two terms on the right-hand side exactly coincide with the source in Eq.~(\ref{eq-evol-rho-inf}) with a flipped sign. Thus, we conclude that these terms describe the energy transfer between the inflaton and gauge fields due to the kinetic and axial couplings  (note that the coupling functions $I_{1,2}$ are explicitly present in those terms and, as expected, the terms vanish if $I_1$ and $I_2$ are constant). 

There is, however, an additional term on the right-hand side of Eq.~(\ref{eq-evol-rho-GF}), namely, the boundary term
\begin{equation}
    [\dot{\rho}_{\mathrm{GF}}]_\mathrm{b}=\frac{1}{2}\Big([\dot{\mathscr{E}}^{(0)}]_\mathrm{b}+[\dot{\mathscr{B}}^{(0)}]_\mathrm{b}\Big).
\end{equation}
It also acts as a source in the energy-conservation equation and describes the energy increase due to contributions of new modes crossing the horizon during inflation. The energy of subhorizon Fourier modes is not included in $\mathscr{E}^{(0)}$ and $\mathscr{B}^{(0)}$ as these modes are considered as vacuum gauge-field fluctuations. After crossing the horizon, these modes must be taken into account and since their energy is nonzero at the moment of horizon crossing, the total energy increases. Thus, the last term in Eq.~(\ref{eq-evol-rho-GF}) can be considered as a vacuum source term. There is no analogous term coming from the inflaton field because we consider it as a spatially homogeneous classical field. If we would take into account vacuum fluctuations of the inflaton above its mean value, a similar term would appear also in Eq.~(\ref{eq-evol-rho-inf}).
 
For practical applications, one has to truncate the infinite system of Eqs.~(\ref{Friedmann-fin})--(\ref{eq-BB1}) at some finite order. Like it was shown previously in the cases of purely kinetic coupling \cite{Sobol:2020lec} or purely axion coupling \cite{Gorbar:2021rlt}, the simplest assumption that one may adopt is the following:
 \begin{eqnarray}
 \label{truncation}
  \mathscr{E}^{(n+1)}&\simeq&\Big(\frac{k_{\mathrm{h}}}{a}\Big)^{2}\mathscr{E}^{(n-1)},\quad \mathscr{B}^{(n+1)}\simeq\Big(\frac{k_{\mathrm{h}}}{a}\Big)^{2}\mathscr{B}^{(n-1)},\nonumber\\
  \mathscr{G}^{(n+1)}&\simeq&\Big(\frac{k_{\mathrm{h}}}{a}\Big)^{2}\mathscr{G}^{(n-1)}.
 \end{eqnarray}
These relations are very natural, especially for large orders $n\gg 1$. First, the symmetry properties under parity transformation of quantities on both sides are the same; i.e., the scalar is related to a scalar, a pseudoscalar\,---\,to pseudoscalar. Second, for any smooth momentum dependence of the mode function, at sufficiently large order $n$ the integral over the spectrum will be dominated by the region close to $k=k_{\mathrm{h}}$. Then, the truncation rule~(\ref{truncation}) follows from the fact that the $(n+1)$th- and $(n-1)$th-order quantities differ  by a factor $(k/a)^2$ under the integral. Thus, the last three equations for the bilinear functions of order $n_{\mathrm{max}}$ differ from all preceding orders and have the form:
\begin{multline}
\dot{\mathscr{E}}^{(n_{\mathrm{max}})}+(n_{\mathrm{max}}+4)H\mathscr{E}^{(n_{\mathrm{max}})}+\frac{\dot{I}_1}{I_1}\mathscr{E}^{(n_{\mathrm{max}})}\\
-2\frac{\dot{I_2}}{I_1}\mathscr{G}^{(n_{\mathrm{max}})}+2\Big(\frac{k_{\mathrm{h}}}{a}\Big)^2\mathscr{G}^{(n_{\mathrm{max}}-1)}=
[\dot{\mathscr{E}}^{(n_{\mathrm{max}})}]_{\mathrm{b}},
\label{eq-EE1-last}
\end{multline}
\begin{multline}
\dot{\mathscr{G}}^{(n_{\mathrm{max}})}+(n_{\mathrm{max}}+4)H\mathscr{G}^{(n_{\mathrm{max}})}
-\frac{\dot{I_2}}{I_1}\mathscr{B}^{(n_{\mathrm{max}})}\\
+\Big(\frac{k_{\mathrm{h}}}{a}\Big)^2[\mathscr{B}^{(n_{\mathrm{max}}-1)}-\mathscr{E}^{(n_{\mathrm{max}}-1)}]=
 [\dot{\mathscr{G}}^{(n_{\mathrm{max}})}]_{\mathrm{b}}, \label{eq-EB1-last}
 \end{multline}
 \begin{multline}
\dot{\mathscr{B}}^{(n_{\mathrm{max}})} + (n_{\mathrm{max}}+4)H\,	\mathscr{B}^{(n_{\mathrm{max}})}-\frac{\dot{I_1}}{I_1}\mathscr{B}^{(n_{\mathrm{max}})}\\
-2\Big(\frac{k_{\mathrm{h}}}{a}\Big)^2\mathscr{G}^{(n_{\mathrm{max}}-1)}=
[\dot{\mathscr{B}}^{(n_{\mathrm{max}})}]_{\mathrm{b}}. \label{eq-BB1-last}
\end{multline}
In Sec.~\ref{sec-numerical} we will see that one can choose the truncation order $n_\mathrm{max}$ in such a way that the solution of the system of equations becomes independent on $n_\mathrm{max}$ itself.

\section{Extended Starobinsky model with nonminimally coupled gauge field}
\label{sec-nonminimal}

In this section we apply the gradient expansion formalism developed above to a specific magnetogenesis model with nonminimally coupled gauge field. As we discussed in the introduction, we are interested in an ``economical'' model which (i)~does not contain any additional scalar fields playing the role of the inflaton and (ii)~has the gauge-field coupling in a simple form with a finite number of free parameters. Both criteria are satisfied for the extended Starobinsky model described below.

\subsection{Nonminimal coupling in Jordan frame}

Let us consider the action, which represents an extension of the Starobinsky inflationary model~\cite{Starobinsky:1980}. Like the Starobinsky model, this action contains a term quadratic in the spacetime curvature $R$; however, the coefficient in front of this term now depends on the gauge field $A_\mu$ (only through the gauge-invariant quantities $F_{\mu\nu}$ and $\tilde{F}_{\mu\nu}$): 
\begin{multline}
\label{action-ext-St}
    S[g_{\mu\nu}, A_\mu]=\int d^4x \sqrt{-g}\\
    \times\Big[-\frac{M_{\mathrm{P}}^2}{2}R+\frac{\xi_s}{4 \Delta(F_{\mu\nu})}R^2-\frac{1}{4}F_{\mu\nu}F^{\mu\nu}\Big],
\end{multline}
where
\begin{equation}
    \Delta(F_{\mu\nu})=1+\frac{\kappa_{1}}{M_{\mathrm{P}}^4}F_{\mu\nu}F^{\mu\nu}+\frac{\kappa_{2}}{M_{\mathrm{P}}^4}F_{\mu\nu}\tilde{F}^{\mu\nu}.
\end{equation}
In this action, $\xi_s$ is a free parameter which can be constrained from the requirement that it leads to the correct amplitude of the scalar power spectrum measured by Planck Collaboration~\cite{Planck:2018-infl}, and the dimensionless coupling constanta $\kappa_{1}$ and $\kappa_{2}$ characterize the nonminimal coupling of the gauge field to curvature which preserves ($\kappa_1$) and which violates ($\kappa_2$)  parity symmetry
($F_{\mu\nu}F^{\mu\nu}$ is the standard Maxwell term which is a scalar quantity, therefore, it preserves parity, while $F_{\mu\nu}\tilde{F}^{\mu\nu}$ is odd under parity transformation, therefore, it is a pseudoscalar quantity).

This model belongs to the class of modified gravity theories and is inspired by the $f(R,\,L_{m})$ \cite{Harko:2010} and Born-Infeld type \cite{BeltranJimenez:2017} theories (however, it certainly does not belong to any of them). Although $\Delta(F_{\mu\nu})$ in Eq.~(\ref{action-ext-St}) does not seem to be positive definite and may vanish at some point causing a singularity in the Lagrangian, we show below  that this never happens during inflation because backreaction of the generated gauge fields never allows the second and third terms to be sufficiently large to cancel the first term (unity).

The specific form of the gauge-field dependence of the $R^2$-term in the Jordan frame was constructed in such a way that in the Einstein frame the gauge-field Lagrangian is quadratic in $A_\mu$ and no higher powers appear (we show this explicitly in the next subsection). This fact allows us to use the gradient expansion formalism which was derived specifically for this type of gauge-field Lagrangian. Moreover, since no higher order terms have been neglected, it is possible to treat the case of a strong gauge field and take into account 
 backreaction on the background evolution. This was impossible to do in our previous work~\cite{Durrer:2022emo} where the form of nonminimal coupling in the Jordan frame $\propto R^n F_{\mu\nu}F^{\mu\nu}$ and $\propto R^n F_{\mu\nu}\tilde{F}^{\mu\nu}$ leads to the presence of terms in the Lagrangian with higher power of the gauge field (in addition to usual quadratic terms). As a result, we could only to study the perturbative regime where the gauge field does not cause backreaction.

Therefore, the proposed model is constructed in such a way that it allows to solve simultaneously several issues. First, similarly to the usual Starobinsky model, its gravity sector contains an additional scalar degree of freedom which plays the role of the inflaton and has an asymptotically flat potential favored by  CMB observations~\cite{Planck:2018-infl}. Second, it can be rewritten in the Einstein frame with a Lagrangian which is quadratic in the gauge field and has kinetic and axial couplings to the inflaton. Finally, due to the absence of higher order terms in the gauge fields, it allows us to study the gauge field nonperturbatively and take into account its backreaction. We will see all these features below.

\subsection{Action in the Einstein frame}
 			
In order to bring action~(\ref{action-ext-St}) to the canonical Einstein-Hilbert form, we  perform a two-step procedure. First, we get rid of the $R^2$ term by performing a Legendre transform and introducing an additional scalar degree of freedom. Second, we perform a conformal transformation of the metric.
	
Let us rewrite the action in the following form:
\begin{equation}
    S[g_{\mu\nu},\,A_{\mu}]=\int d^4x \sqrt{-g}\Big[-\frac{M_{\mathrm{P}}^2}{2} f(R,F_{\mu\nu})-\frac{1}{4}F_{\mu\nu}F^{\mu\nu}\Big],
\end{equation}
where
\begin{equation}
    f(R,\,F_{\mu\nu})=R-\frac{\xi_s}{2M_{\mathrm{P}}^{2}\Delta(F_{\mu\nu})}R^2.
\end{equation}
We now introduce the new auxiliary field
\begin{equation}
    \Psi=\frac{\partial f}{\partial R}=1-\frac{\xi_s}{M_{\mathrm{P}}^{2}\Delta(F_{\mu\nu})}R
\end{equation}
and express the spacetime curvature as
\begin{equation}
    R=\frac{M_{\mathrm{P}}^{2}}{\xi_{s}}\Delta(F_{\mu\nu})\big(1-\Psi\big).
\end{equation}
Then, the Legendre transform of the function $f$ reads 
\begin{equation}
    F(\Psi,\, F_{\mu\nu})=\Psi R-f(R,\,F_{\mu\nu})= -\frac{M_{\mathrm{P}}^{2}}{2\xi_{s}} \Delta(F_{\mu\nu})\big(1-\Psi\big)^{2}.
\end{equation}
Finally, representing $f$ as an inverse Legendre transform $f(R,\, F_{\mu\nu})=\Psi R -F(\Psi,\, F_{\mu\nu})$, we obtain the action in the form
\begin{multline}
    S[g_{\mu\nu},\,\Psi,\, A_{\mu}]=\int d^4x \sqrt{-g}\Big[-\frac{M_{\mathrm{P}}^2}{2}\Psi R\\
    -\frac{M_{\mathrm{P}}^4}{4\xi_s}\Delta(F_{\mu\nu}) \big(1-\Psi\big)^{2} 
    -\frac{1}{4}F_{\mu\nu}F^{\mu\nu}\Big],
\end{multline}
which is linear in $R$ and contains an additional scalar degree of freedom, $\Psi$ ( as the action does not contain a kinetic term for it, $\Psi$ is nondynamical). Note, that already now the function $\Delta(F_{\mu\nu})$ appears only in the numerator and the overall action is quadratic in the gauge field.
	
In order to remove the extra multiplier $\Psi$ in front of $R$, we perform the Weyl transformation $g_{\mu\nu}= \Psi^{-1} \bar{g}_{\mu\nu}$, under which the Ricci curvature scalar transforms as
\begin{equation}
    R=\Psi \Big[\bar{R}-\frac{3}{2\Psi^{2}}\bar{g}^{\mu\nu}\partial_{\mu}\Psi\partial_{\nu}\Psi+3\bar{\nabla}_{\mu}\bar{\nabla}^{\mu}\ln\Psi\Big].
\end{equation}
This brings us to the Einstein frame where the action takes the form
\begin{multline}
\label{action-ext-St-Psi}
    S[\bar{g}_{\mu\nu},\,\Psi,\, A_{\mu}]=
    \int d^4x \sqrt{-\bar{g}}\Big[-\frac{M_{\mathrm{P}}^2}{2}\bar{R}
    +\frac{3M_{\mathrm{P}}^2}{4\Psi^2}\partial_\mu \Psi\partial^\mu \Psi
    \\-\frac{M_{\mathrm{P}}^4}{4\xi_s}\frac{(1-\Psi)^2}{\Psi^2}\Delta(F_{\mu\nu})
    -\frac{1}{4}F_{\mu\nu}F^{\mu\nu}\Big].
\end{multline}
Here the function $\Delta$ reads as
\begin{equation}
\label{Delta-new}
    \Delta(F_{\mu\nu})=1+\frac{\kappa_{1}}{M_{\mathrm{P}}^4}\Psi^2 F_{\mu\nu}F^{\mu\nu}+\frac{\kappa_{2}}{M_{\mathrm{P}}^4}\Psi^2 F_{\mu\nu}\tilde{F}^{\mu\nu}
\end{equation}
and all contractions of indices are now performed by means of the new metric $\bar{g}_{\mu\nu}$. In action~(\ref{action-ext-St-Psi}) the field $\Psi$ is dynamical, however, it is not canonically normalized. In order to fix this, we perform the following field redefinition
\begin{equation}
\label{Psi}
    \Psi=\exp\Big(\sqrt{\frac{2}{3}}\frac{\phi}{M_{\mathrm{P}}}\Big),
\end{equation}
which finally leads us to 
\begin{equation}
    S[\bar{g}_{\mu\nu},\,\phi,\, A_{\mu}]=S_{EH}[\bar{g}_{\mu\nu}] + S_{\mathrm{KA}}[\bar{g}_{\mu\nu},\,\phi,\, A_{\mu}],
\end{equation}
where the first term is the Einstein-Hilbert action for gravity and the second term 
is the exactly as given in Eq.~(\ref{action-kinetic-axial}) and describes the inflaton and gauge fields with kinetic and axial couplings. The potential of the inflaton field $V(\phi)$ coincides with corresponding expression for the Starobinsky model 
\begin{equation}
\label{potential-general}
    V(\phi)=\frac{M_{\mathrm{P}}^4}{4 \xi_s}\Big(1-e^{-\sqrt{\frac{2}{3}}\frac{\phi}{M_{\mathrm{P}}}}\Big)^2
\end{equation}
while the kinetic and axial coupling functions have the  form
\begin{equation}
\label{I1}
    I_{j}=\delta_{j1}+\frac{\kappa_j}{\xi_s}\left[\exp\left(\sqrt{\frac{2}{3}}\frac{\phi}{M_{\mathrm{P}}}\right)-1\right]^2.
\end{equation}
Note that the dependence of these coupling functions on the inflaton field is not postulated or constructed by hand, but is deduced from a simple action~(\ref{action-ext-St}) by rewriting it in the Einstein frame. The only freedom, which is left, is contained in the two dimensionless parameters $\kappa_1$ and $\kappa_2$. The general requirements for the kinetic coupling function [listed in a paragraph below Eq.~(\ref{action-kinetic-axial})] are automatically satisfied by $I_1$ in Eq.~(\ref{I1}) if $\kappa_1 >0$. We will consider different values of these parameters in the next section devoted to a numerical analysis. 

A few additional comments are in order. First of all, we want to emphasize that the action in the Jordan frame, Eq.~(\ref{action-ext-St}), was intentionally chosen  such  that  we obtain the flat Starobinsky potential for the inflaton and a quadratic Lagrangian for the gauge field in Einstein frame. The former fact is important as we want an inflationary model which is in agreement with recent CMB observations~\cite{Planck:2018-infl}, while the latter condition allows us to study also a nonperturbative strong-field regime when backreaction of the produced field significantly changes the background evolution.

Another interesting observation is that the coupling function in Eq.~(\ref{I1}) has exactly the same form as in the model of Higgs or Higgs-Starobinsky inflation (in the metric formulation) with nonminimally coupled gauge field through the terms $\propto R^2 F_{\mu\nu}F^{\mu\nu}$ and $\propto R^2 F_{\mu\nu}\tilde{F}^{\mu\nu}$ considered by us in Ref.~\cite{Durrer:2022emo}. In that work we studied only the perturbative regime where  backreaction does not occur. The study of the backreaction regime in that framework would be not self-consistent because the higher-order terms were omitted in the derivation of the gauge-field action in Einstein frame. The model we chose in the present article does not have such a restriction and, thus, our present results may qualitatively describe the onset of the backreaction regime also in the model in Ref.~\cite{Durrer:2022emo}.

Finally, before performing a numerical analysis, we would like to make sure that the function $\Delta(F_{\mu\nu})$ in the denominator in Eq.~(\ref{action-ext-St}) is always positive during inflation. Indeed, let us consider the vacuum expectation value of Eq.~(\ref{Delta-new}) and express it in terms of bilinear quantities (\ref{EE})--(\ref{BB}):
\begin{equation}
    \langle \Delta(F_{\mu\nu})\rangle = 1+\frac{2\Psi^2}{M_{\mathrm{P}}^4 I_{1}}\Big[\kappa_1 (\mathscr{B}^{(0)}-\mathscr{E}^{(0)})+ 2\kappa_2 \mathscr{G}^{(0)}\Big].
\end{equation}
Using the explicit form of the coupling functions~(\ref{I1}), the potential~(\ref{potential-general}) and Eq.~(\ref{Psi}), it is straightforward to check that
\begin{equation}
    \langle \Delta(F_{\mu\nu})\rangle = 1-e^{-\sqrt{\frac{2}{3}}\frac{\phi}{M_{\mathrm{P}}}}\frac{\frac{I_1^{\prime}(\phi)}{2I_1} (\mathscr{E}^{(0)}-\mathscr{B}^{(0)})-\frac{I_2^{\prime}(\phi)}{I_1}  \mathscr{G}^{(0)}}{V'(\phi)},
\end{equation}
where the expression in the numerator of the long fraction is the right-hand side of the Klein-Gordon equation~(\ref{KGF-3}), while the denominator is the ``drag force'' term on the left-hand side of the same equation. Note that a prime denotes a derivative with respect to the
inflaton. It is convenient to introduce a dimensionless parameter which quantifies the relevance of the backreaction term in the Klein-Gordon equation, $\delta_{\mathrm{BR}}$, see Eq.~(\ref{delta-BR}) below. Then, we may estimate the value of $\langle\Delta\rangle$ as follows:
\begin{equation}
\label{Delta-fin}
    \langle \Delta(F_{\mu\nu})\rangle\geq 1-e^{-\sqrt{\frac{2}{3}}\frac{\phi}{M_{\mathrm{P}}}} \delta_{\mathrm{BR}} >0.
\end{equation}
Indeed, as we will show in the next section, in the strong backreaction regime $\delta_{\mathrm{BR}}\simeq 1$.\footnote{In the case of purely axial coupling the corresponding solution for the inflaton field in the backreaction regime is often referred to as the Anber-Sorbo attractor since it was studied for the first time in Ref.~\cite{Anber:2010}; see the recent article~\cite{Peloso:2022} for the analysis of (in)stability of this attractor.} However, in Eq.~(\ref{Delta-fin}), it is multiplied by a factor $e^{-\sqrt{\frac{2}{3}}\frac{\phi}{M_{\mathrm{P}}}}$ which is exponentially small far from the end of inflation. On the contrary, close to the end of inflation $e^{-\sqrt{\frac{2}{3}}\frac{\phi}{M_{\mathrm{P}}}}\lesssim 1$ while $\delta_{\mathrm{BR}}\ll 1$, because the produced gauge field typically decreases toward the end of inflation. Therefore, the product in the second term is always smaller than unity and there is no danger that $\langle \Delta(F_{\mu\nu})\rangle$ would vanish during inflation.

\section{Numerical results}
\label{sec-numerical}

\subsection{General remarks}

In this section we perform a numerical analysis of gauge-field production in the extended Starobinsky model introduced in the previous section. First, we specify the numerical values of our model parameters. As we discussed earlier, the parameter $\xi_s$ which determines the magnitude of the $R^2$ term in action~(\ref{action-ext-St}) in Jordan frame and the amplitude of the Starobinsky potential~(\ref{potential-general}) can be fixed from CMB observations because it also determines the amplitude of the scalar power spectrum, $A_s$. For the Starobinsky potential, we have
\begin{equation}
    A_s=\frac{H^2}{8\pi^2 M_{\mathrm{P}}^2 \epsilon}\bigg|_{N_{\ast}}\simeq \frac{N_{\ast}^{2}}{72\pi^2 \xi_s},
\end{equation}
where $\epsilon=\frac{M^2_{\mathrm{P}}}{2}
\big(\frac{V'(\phi)}{V(\phi)}\big)^2$ is the first slow-roll parameter and $N_{\ast}$ is the number of $e$-foldings before the end of inflation when the scalar perturbation mode of the CMB pivot scale $k_{\ast}/a_0=0.05\,\text{Mpc}^{-1}$ exits the horizon. Depending on the details of reheating stage after inflation, $N_{\ast}$ typically lies in the range 50--60. Taking for definiteness $N_{\ast}=50$, and using the best-fit value of the parameter $A_s$ from the Planck 2018 data~\cite{Planck:2018-infl}, $A_s=2.10\times 10^{-9}$, we obtain
\begin{equation}\label{e:xisnum}
    \xi_s=1.68\times 10^9
\end{equation}
and the amplitude of the potential $V_0=M_{\mathrm{P}}^4/(4\xi_s)\simeq 1.5\times 10^{-10}M_{\mathrm{P}}^4$. Even though the backreaction of the produced gauge field may extend the inflation stage, for definiteness, we will still use this normalization of the potential in all our numerical computations.

In order to describe the gauge field production we employ the gradient-expansion formalism introduced in Sec.~\ref{sec-gef}. For this we need to choose the optimal truncation order $n_{\mathrm{max}}$. This is done by increasing the value of $n_{\mathrm{max}}$ until the numerical result for the generated energy densities becomes independent of $n_{\mathrm{max}}$ (further increase of $n_{\mathrm{max}}$ does not change the result). Initial conditions for the inflaton are taken at the slow-roll attractor: the initial value for the inflaton $\phi(0)=\phi_0$ is chosen to provide at least 60-70 $e$-foldings of inflation while the derivative $\dot{\phi}(0)$ is determined as
\begin{equation}
    \dot{\phi}(0)=-\frac{V'(\phi_0)}{3H}\simeq 
    -\frac{M_{\mathrm{P}}^2}{3}\sqrt{\frac{2}{\xi_s}}e^{-\sqrt{\frac{2}{3}}\frac{\phi_0}{M_{\mathrm{P}}}}.
\end{equation}
Concerning initial conditions for the bilinear functions $\mathscr{E}^{(n)}$, $\mathscr{G}^{(n)}$, and $\mathscr{B}^{(n)}$, they may be chosen to be zero or some nonzero values following from the contributions of vacuum fluctuations outside the horizon at the initial moment of time. The system quickly ``forgets'' these initial conditions and the final result does not depend on them.
More details on the numerical procedure that we apply, additional approximations that we do in the gradient-expansion formalism, and the analysis of the accuracy of the gradient-expansion result are discussed in Appendix~\ref{app-B}.

The parameters $\kappa_1$ and $\kappa_2$, which determine the kinetic and axion coupling of the gauge field to the inflaton, respectively, always enter the coupling functions in the combination $\kappa_{1,2}/\xi_s$. Thus, this combinations play the role of coupling constants and therefore we will specify the numerical values of these combinations rather than the values of $\kappa_{1,2}$. However, one can easily infer the l$\kappa_{1,2}$ using \eqref{e:xisnum} for  $\xi_s$. In the three subsections below we study different cases of interplay between the kinetic and axial couplings. Therefore, we specify the values of $\kappa_{1,2}/\xi_s$, which are used in the computations, directly in the corresponding subsections. 

As we already discussed above, $\kappa_1$ can only be positive in order to avoid the strong coupling problem during inflation, while the sign of $\kappa_2$ is not fixed. However, for definiteness, we will also take it positive. In this case the parameter $\xi(t)$ defined in Eq.~(\ref{xi-s}) is negative and the mode with left circular polarization ($\lambda =-1$) undergoes tachyonic instability and becomes enhanced. Changing the sign of $\kappa_2$ would lead to the enhancement of the right-handed circular polarization and, thus, switch the sign of all bilinear functions $\mathscr{E}^{(2n+1)}$, $\mathscr{G}^{(2n)}$, and $\mathscr{B}^{(2n+1)}$. This situation can be obtained from the case with positive $\kappa_2$ by means of a parity transformation; therefore, it suffices to study only one sign of $\kappa_2$.
 
We have  already noted that the backreaction may affect both the Friedmann equation
(\ref{Friedmann-fin}) through the energy density of produced gauge fields and the inflaton equation~(\ref{KGF-3}) of motion due to a nontrivial source term on the right-hand side. The first effect can be characterized by the following dimensionless parameter
\begin{equation}
    \delta_{\mathrm{ED}}=\frac{\rho_{\mathrm{GF}}}{\rho_{\mathrm{inf}}}=\frac{\frac{1}{2}(\mathscr{E}^{(0)}+\mathscr{B}^{(0)})}{\frac{1}{2}\dot{\phi}^2+V(\phi)}.\label{delta-ED}
\end{equation}
Since in the inflationary regime (accelerated expansion of the Universe), the effective equation of state  $p=w\rho$ must satisfy $w\leq -1/3$, this puts a constraint on the gauge-field energy density which can be achieved during inflation, namely,
\begin{multline}
    w=\frac{p}{\rho}=\frac{\frac{1}{2}\dot{\phi}^2-V(\phi)+\frac{1}{3}\rho_{\mathrm{GF}}}{\frac{1}{2}\dot{\phi}^2+V(\phi)+\rho_{\mathrm{GF}}}\leq -\frac{1}{3} 
    \\ \Rightarrow \quad \rho_{\mathrm{GF}}\leq V(\phi)-\dot{\phi}^2.
\end{multline}
Taking this into account, we conclude that during inflation $\delta_{\mathrm{ED}}\leq -2+(3V/\rho_{\mathrm{inf}})<1$. Nevertheless, its values of order unity would mean that the gauge-field energy density has already a tangible effect on the background expansion.

In order to characterize the importance of different terms in the Klein-Gordon equation~(\ref{KGF-3}), we introduce the following parameters:
\begin{eqnarray}
    \delta_{\mathrm{acc}}&=&\bigg|\frac{\ddot{\phi}}{V'(\phi)}\bigg|,\label{delta-acc}\\
    \delta_{\mathrm{SR}}&=&\bigg|\frac{3H\dot{\phi}}{V'(\phi)}\bigg|,\label{delta-SR}\\
    \delta_{\mathrm{BR}}&=&\bigg|\frac{I_{1}^{\prime}(\phi)(\mathscr{E}^{(0)}-\mathscr{B}^{(0)}) -2 I_2^{\prime}(\phi)\mathscr{G}^{(0)}}{2I_{1}(\phi) V'(\phi)}\bigg|.\label{delta-BR}
\end{eqnarray}
The first of them, $\delta_{\mathrm{acc}}$, quantifies the contribution of the inflaton acceleration, the second one, $\delta_{\mathrm{SR}}$, allows us to conclude whether the slow-roll attractor is realized (if $\delta_{\mathrm{SR}}\simeq 1$), while the third one, $\delta_{\mathrm{BR}}$, tells about the importance of the backreaction of generated gauge fields on the inflaton evolution. In the following subsections we use all four parameters to analyze different stages of evolution of the coupled system ``inflaton+gauge field.''

\subsection{Axial-dominated limit $\kappa_{2}\gg \kappa_{1}$}
\label{subsec-axial}

We start with the axial-dominated case when $\kappa_{2}\gg \kappa_{1}$.  In our previous work \cite{Durrer:2022emo} where a similar type of coupling between the inflaton and gauge field was treated perturbatively, we showed that for the coupling function (\ref{I1}) the generated gauge field typically decreases in time in the axially-dominated limit $\kappa_{2}\gg \kappa_{1}$. Therefore, the backreaction may occur only before a certain moment during inflation and after that the usual slow-roll regime is restored.

For definiteness, we choose the gauge-field coupling parameters as $\kappa_{1}/\xi_s=10^{-5}$, $\kappa_{2}/\xi_{s}=0.15$. In this case,  backreaction occurs at about 20 $e$-foldings before the end of inflation. In order to study this regime numerically, we employ the gradient expansion formalism with truncation order in a range $n_{\mathrm{max}}=150-200$.

\begin{figure*}[ht]
\centering
\includegraphics[width=0.46\linewidth]{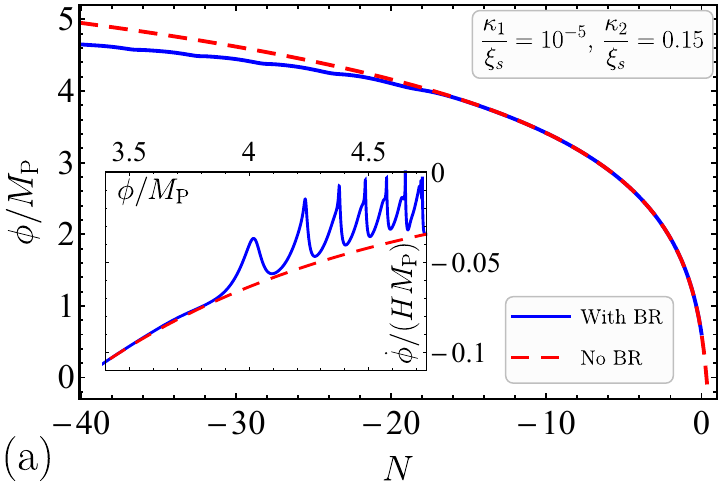}
\hspace{3mm}
\includegraphics[width=0.49\linewidth]{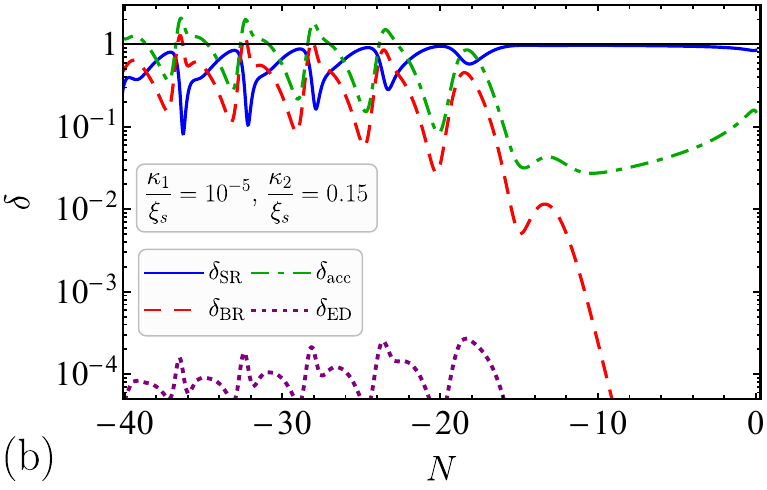}
\caption{(a) The dependence of the inflaton field $\phi$ on the number of $e$-foldings counted from the end of inflation in the absence of gauge fields (red dashed line) and for the case $\kappa_{1}/\xi_s=10^{-5}$, $\kappa_{2}/\xi_{s}=0.15$ where the backreaction occurs (blue solid line). The inset shows a part of the phase portrait of the inflaton field in dimensionless coordinates ($\phi/M_{\mathrm{Pl}},\,\dot{\phi}/(HM_{\mathrm{Pl}})$). (b) The $\delta$-parameters defined in Eq.~(\ref{delta-ED}), (\ref{delta-acc})--(\ref{delta-BR}) as functions of the number of $e$-foldings: the slow-roll parameter $\delta_{\mathrm{SR}}$ is shown by the blue solid line, the backreaction parameter $\delta_{\mathrm{BR}}$\,---\,by the red dashed line, the inflaton acceleration parameter $\delta_{\mathrm{acc}}$\,---\,by the green dashed-dotted line, and the energy-density parameter $\delta_{\mathrm{ED}}$\,---\,by the purple dotted line.}
\label{fig-inflaton-AD}
\end{figure*}

\begin{figure*}[ht]
\centering
\includegraphics[width=0.99\linewidth]{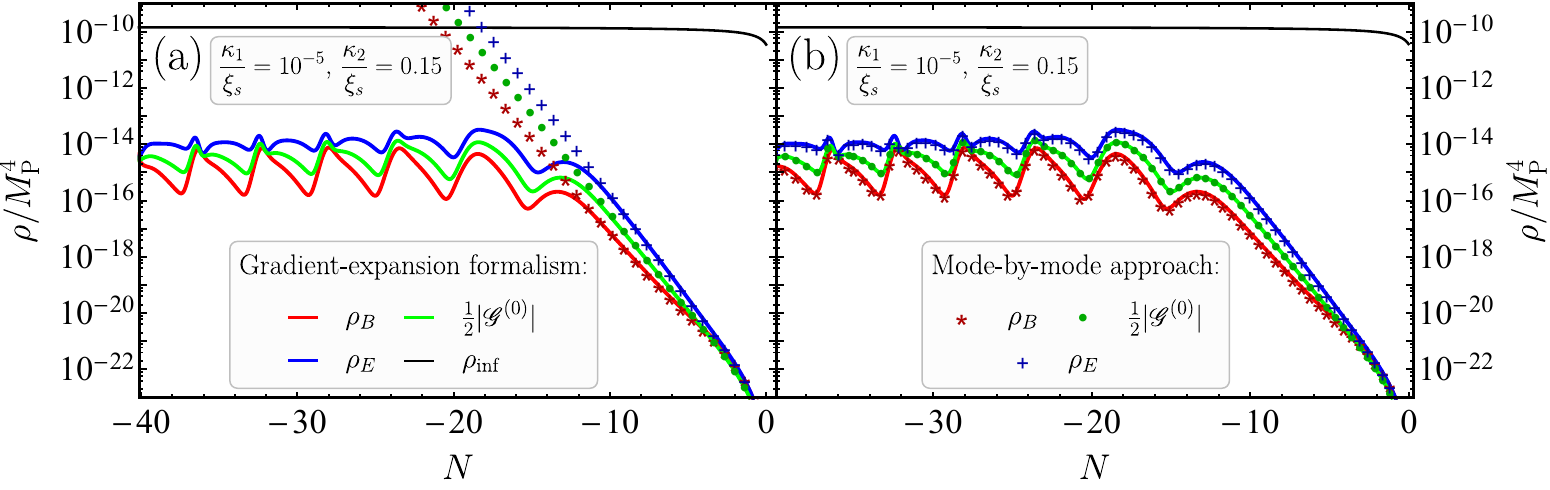}
\caption{The dependence of the energy densities on the number of $e$-foldings counted from the end of inflation in the case $\kappa_{1}/\xi_s=10^{-5}$, $\kappa_{2}/\xi_{s}=0.15$: magnetic energy density (red line), electric energy density (blue line), Chern-Pontryagin density $\mathscr{G}^{(0)}$ (green line), and the energy density of the inflaton (black line). Dots marked by symbols show the corresponding quantities computed in the mode-by-mode approach: (a) when the backreaction is not taken into account, (b) when the backreaction modifies the inflaton evolution (based on the inflaton time dependence taken from the result of the gradient-expansion formalism).}
\label{fig-rho-AD}
\end{figure*}

There are a few features in the backreaction regime that should be noted.
First, the inflaton field rolls more slowly than in the absence of
backreaction of generated gauge fields; e.g., in a particular case shown in Fig.~\ref{fig-inflaton-AD}(a), in the presence of backreaction, the inflaton field at any number of $e$-foldings before the end of inflation is always smaller than in the absence of backreaction. In other words, in order to roll down from the same initial value of the inflaton field, it takes 10--15 $e$-foldings more in the presence of backreaction. The same conclusion follows from the analysis of the inflaton phase portrait which is shown in the inset in Fig.~\ref{fig-inflaton-AD}(a). Indeed, in the backreaction regime (blue solid line), the inflaton velocity is always smaller in absolute value than in the absence of gauge fields (red dashed line).

Second, Figs.~\ref{fig-inflaton-AD}(b) and \ref{fig-rho-AD} clearly demonstrate that the energy density condition $\rho_\mathrm{GF}\ll \rho_\mathrm{inf}$ is always satisfied even in the backreaction regime, meaning that $\delta_{\mathrm{ED}} \sim 1$ is not a necessary condition for the occurrence of backreaction. The presence of the gauge field in the equation of motion for the inflaton field,  Eq.~(\ref{KGF-3}), appears to be more important. Indeed, the parameter $\delta_{\mathrm{BR}}$ [shown by the red dashed line in Fig.~\ref{fig-inflaton-AD}(b)] is of order unity all the time up to 20 $e$-foldings before the end of inflation. After that the inflaton enters the usual slow-roll regime, see the blue solid curve for $\delta_{\mathrm{SR}}$ which tends to unity and the green dashed-dotted line for $\delta_{\mathrm{acc}}$ which becomes much less than unity. Thus, in order to track  backreaction effects, one needs first to consider the parameter $\delta_{\mathrm{BR}}$.

Third, the behavior of the generated energy densities as well as the inflaton field in the backreaction regime are not monotonic, but an oscillatory behavior is observed. This can be explained by the retardation of the gauge-field response to the changes of the inflaton field~\cite{Domcke:2020}: gauge-field modes which are being enhanced at a certain moment of time will become dominant in the energy density and $\mathscr{G}^{(0)}$ only after some retardation time of a few $e$-foldings. We discuss these oscillations in more details in Sec.~\ref{subsec-oscillations}.

Finally, we want to point out that the widely used assumption that in the backreaction regime the inflaton experiences a different form of a slow-roll behavior where the drag force from the inflaton potential is almost exactly compensated by the backreaction term on the right-hand side of Eq.~(\ref{KGF-3}) \cite{Anber:2010} is not well justified. Indeed, the two other terms, the inflaton acceleration and the Hubble friction, appear to be of the same order of magnitude as the potential term, see the red dashed line for $\delta_{\mathrm{SR}}$ and the green dashed-dotted line for $\delta_{\mathrm{acc}}$ in Fig.~\ref{fig-inflaton-AD}(b). They both are oscillating and acquire the values of order unity. Moreover, they show the opposite behavior of  the backreaction parameter $\delta_{\mathrm{BR}}$: when it is maximal, they both are minimal and vice versa; however, all three of them on average are of order unity.

In order to do a consistency check of the numerical results obtained in the gradient expansion formalism, we apply the standard approach in momentum space and solve the mode equation (\ref{eq-mode-2}) for all physically relevant modes. Then, integrating the spectral densities over the momenta of all modes which have been amplified, we obtain the generated gauge-field energy densities. First, the mode equation is solved using the unperturbed time dependence of the inflaton field (i.e., neglecting the backreaction of the generated fields on the background evolution). The results are shown by dotted lines in Fig.~\ref{fig-rho-AD}(a). As we see in this figure, these results are in good accordance with the results obtained using the gradient expansion formalism during the last 10 $e$-foldings of inflation, where the generated fields are weak and backreaction is irrelevant. However, large deviations occur at earlier times. In a second approach, we take time dependence of the inflaton field and scale factor from the numerical solution of the gradient-expansion formalism and use it to solve the mode equation (\ref{eq-mode-2}). The numerical results of this mode-by-mode treatment are shown in Fig.~\ref{fig-rho-AD}(b) by dotted lines. We see a very good agreement between the two results although there is a small residual error (typically less than 1\,\% in the backreaction regime reaching a few percent level during the last 3 $e$-foldings of inflation).
This analysis confirms that the gradient-expansion formalism gives adequate results for the generated gauge field in the strong backreaction regime.

\begin{figure}[ht]
\centering
\includegraphics[width=0.97\linewidth]{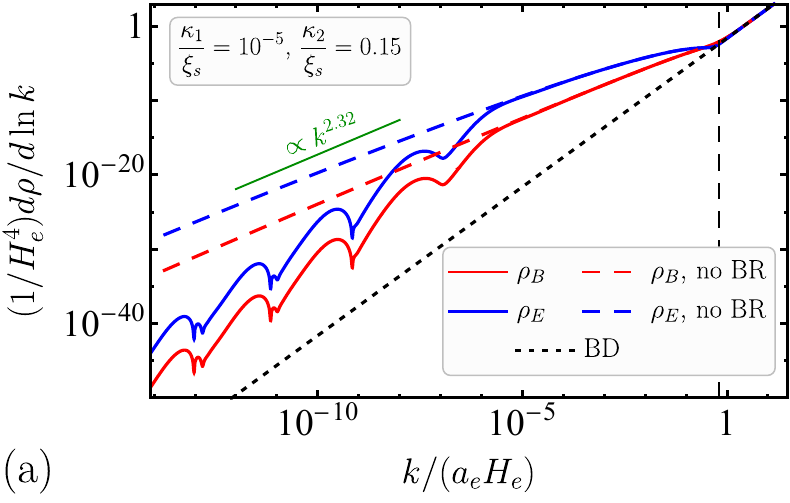}\\
\vspace*{3mm}
\includegraphics[width=0.97\linewidth]{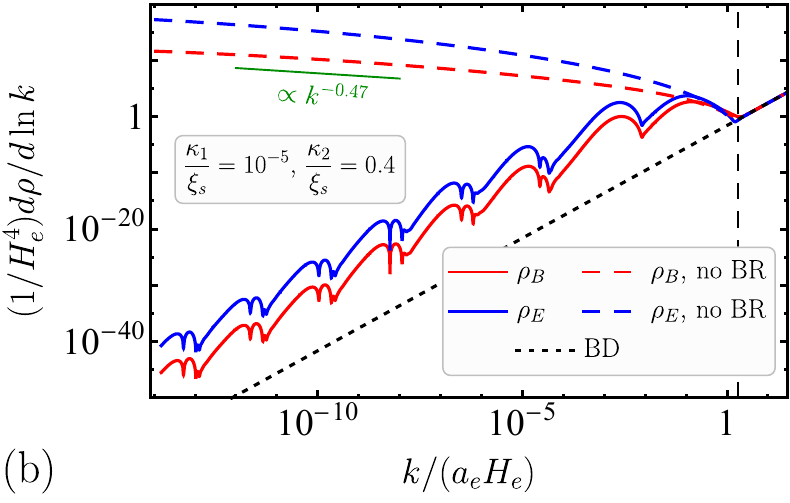}
\caption{The spectra of the generated magnetic (red) and electric (blue) energy densities at the end of inflation for $\kappa_{1}/\xi_s=10^{-5}$ and (a)~$\kappa_{2}/\xi_{s}=0.15$ or (b)~$\kappa_{2}/\xi_{s}=0.4$. The solid lines show the spectra in the case when the backreaction is taken into account while the dashed lines correspond to the case when backreaction is absent. The thin green solid lines correspond to a constant spectral index computed according to Eq.~(\ref{nB-analytic}). \label{fig-spectrum-AD}}
\end{figure}

The mode-by-mode treatment in momentum space also allows to find the spectrum of the generated gauge fields. We show the magnetic and electric power spectra in Fig.~\ref{fig-spectrum-AD} by the red and blue lines, respectively. The dashed lines show the spectra of the mode equation~(\ref{eq-mode-2}) neglecting backreaction while the solid lines correspond to the case when backreaction is taken into account.

We would like to mention that in the case of the gauge-field production without backreaction in this model of kinetic and axial coupling with the coupling function~(\ref{I1}) was studied by us in Ref.~\cite{Durrer:2022emo}. Therefore, in order to analyze the spectra without backreaction, we can use some of the results obtained there. In particular, the magnetic spectral index in the axial-dominated regime reads as
\begin{equation}
\label{nB-analytic}
    n_{B}=4-\frac{32\pi}{9}\frac{\kappa_2}{\xi_s}. 
\end{equation}
Thus, choosing a sufficiently large value of the coupling parameter $\kappa_2/\xi_s > 9/(8\pi)\approx 0.36$, we may get a red-tilted spectrum. In Fig.~\ref{fig-spectrum-AD}, we show the spectra for $\kappa_2/\xi_s=0.15$ [panel (a)] and for $\kappa_2/\xi_s=0.4$ [panel (b)]. In full accordance with our analytical estimates, in the former case the spectrum is always blue-tilted while for the latter value it becomes red-tilted [the thin green lines sketch the tilt of spectrum with $n_B$ given by Eq.~(\ref{nB-analytic})]. However,  backreaction leads to drastic changes of the spectral shape in both cases. First, for all modes which cross the horizon in the backreaction regime the spectrum becomes blue-tilted (on average) and the average spectral tilt is $n_{B}\simeq 4$ independent of the value of the coupling parameter $\kappa_2/\xi_s$. Second, the spectrum also shows oscillatory behavior which reflects the inflaton oscillations in the backreaction regime. Peaks in the power spectrum occur for the modes which cross the horizon at the moment when the inflaton velocity is maximal. 

The final part of the spectrum, for modes which cross the horizon during the last few $e$-foldings of inflation (when there is no backreaction), has a spectral tilt which is close to the value~(\ref{nB-analytic}). This means, that for sufficiently large $\kappa_2/\xi_s$ we can still get a red-tilted spectrum, although for a limited range of modes which correspond to very small scales, see Fig.~\ref{fig-spectrum-AD}(b). This may help to get a larger correlation length of the produced gauge, however, only 2--3 orders of magnitude larger than the horizon size at the end of inflation.

Thus, we conclude that in the axial-dominated case, due to the fact that the coupling function $I_2$ in Eq.~(\ref{I1}) always grows with the increase of $\phi$, we unavoidably face a backreaction problem which is more severe further from the end of inflation. This raises an important issue with the spectra of primordial scalar perturbations. The modes of scalar perturbations which cross the horizon 50--60 $e$-foldings before the end of inflation make an imprint in the CMB anisotropy spectrum. Therefore,  backreaction occurring at this time will definitely modify the predictions of inflationary theory and may even contradict CMB observations. If the coupling parameter is so small that the backreaction is  absent at 50--60 $e$-foldings before  the end of inflation, then the resulting gauge field are negligibly small. Consequently, we need another effect which may turn off backreaction in the time interval when the modes relevant for CMB cross the horizon. Fortunately, this can be easily done by including a relevant kinetic coupling. We study this possibility in the following subsections.

\subsection{Kinetic-dominated limit $\kappa_{1}\gg \kappa_{2}$}
\label{subsec-kinetic}

\begin{figure*}[ht]
\centering
\includegraphics[width=0.99\linewidth]{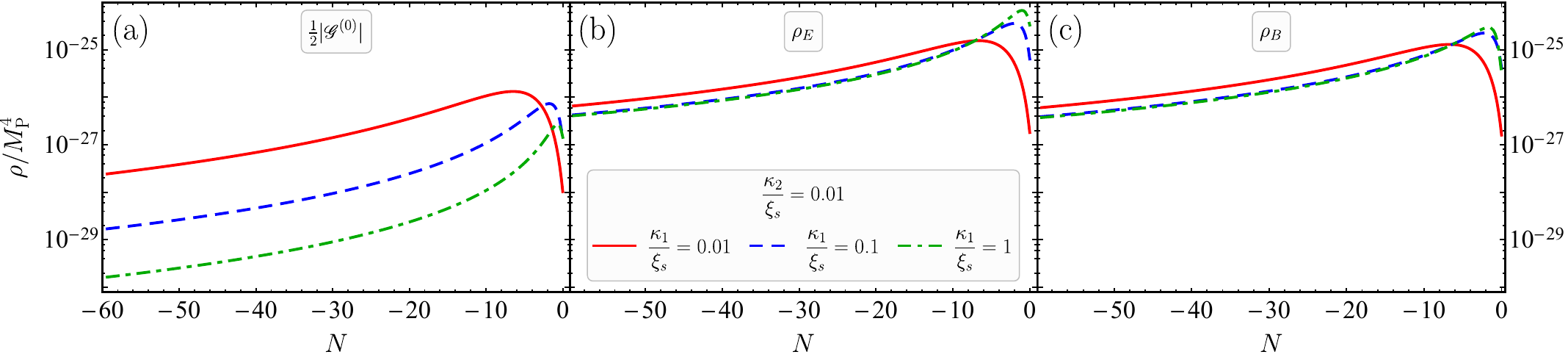}
\caption{The dependence of (a)~the Chern-Pontryagin density $\mathscr{G}^{(0)}$, (b)~the electric energy density, and (c)~the magnetic energy density on the number of $e$-foldings counted from the end of inflation in the case $\kappa_{2}/\xi_s=10^{-2}$ for three values of the parameter $\kappa_1$: $\kappa_{1}/\xi_{s}=0.01$ (red solid lines), $\kappa_{1}/\xi_{s}=0.1$ (blue dashed lines), and $\kappa_{1}/\xi_{s}=1$ (green dashed-dotted lines).}
\label{fig-rho-KD}
\end{figure*}

Another limiting case is the domination of the kinetic coupling when $\kappa_{1}\gg \kappa_{2}$. However, this regime is less interesting as far as magnetogenesis is concerned. Let us discuss its features in more details. First, we note that for a meaningful value of $\kappa_1/\xi_s\sim 10^{-2}-1$, most of the time during inflation the second term in the coupling function~(\ref{I1}) dominates; only during the last few $e$-foldings of inflation, the unity in Eq.~(\ref{I1}) also becomes relevant. Therefore, most of the time $\kappa_1$ cancels everywhere in equations of motion except in the terms with axial coupling function $I_2$ where it remains in the combination $\kappa_2/\kappa_1$. This is because the coupling function $I_1$ enters equations of motion either in the form of $I_1^{\prime}(\phi)/I_1$ or $I_2^{\prime}(\phi)/I_1$. This fact has different consequences for parity-odd and parity-even bilinear gauge-filed functions. If the function is parity-odd, namely $\mathscr{E}^{(2n+1)}$, $\mathscr{B}^{(2n+1)}$, or $\mathscr{G}^{(2n)}$, it is significantly suppressed compared to the parity-even functions of the same order, $\mathscr{G}^{(2n+1)}$, $\mathscr{E}^{(2n)}$, and $\mathscr{B}^{(2n)}$. Indeed, in the spectral representation of parity-odd functions, contributions of modes with left- and right-handed circular polarization enter with opposite signs, see Eqs.~(\ref{E_1})--(\ref{B_1}). The two polarizations evolve differently only because of nonzero axial coupling which is controlled by the ratio $\kappa_2/\kappa_1 \ll 1$. Therefore, for small values of $\kappa_2/\kappa_1$ the two polarizations evolve similar and parity-odd functions are suppressed. At the same time, parity-even functions remain almost insensitive to the value of $\kappa_2/\kappa_1$ and, thus, to any of them. These features are clearly seen in Fig.~\ref{fig-rho-KD}, where panel~(a) shows the parity-odd function $\mathscr{G}^{(0)}$ while panels~(b) and (c) show the parity-even functions $\rho_{E}=\mathscr{E}^{(0)}/2$ and $\rho_B=\mathscr{B}^{(0)}/2$, respectively.

Even if we consider parity-even functions, e.g., the electric and magnetic energy densities, the produced values of these quantities appear to be very small. This can be explained by the fact that the logarithm of the coupling function $I_1$\,---\,the only thing which is important for parity-even quantities\,---\,is a slowly varying function during  slow-roll inflation, $\ln{( I_{1}(\phi))} \simeq 2\sqrt{2/3} (\phi/M_{\mathrm{P}})+\text{const.}$ This problem cannot be resolved by increasing the coupling constant $\kappa_1$, because, as we discussed above, it disappears from equations of motion.

In the last few $e$-foldings of inflation, when the second term in the coupling function~(\ref{I1}) becomes subdominant compared to unity, i.e., for $\phi < M_{\mathrm{P}}\sqrt{3/2}\ln [1+(\kappa_1/\xi_s)^{-1/2}]$, the dependence on $\kappa_1$ recovers. However, the functions $I_j^{\prime}\sim (\kappa_j/\xi_s) e^{\sqrt{\frac{2}{3}}\frac{\phi}{M_{\mathrm{P}}}}(e^{\sqrt{\frac{2}{3}}\frac{\phi}{M_{\mathrm{P}}}}-1)$ are decreasing in time (since $\phi$ decreases) and this leads to the decrease of all bilinear functions (see the final part of the curves in Fig.~\ref{fig-rho-KD}). Thus, we conclude that in kinetic-dominated case, generated gauge fields are always very weak. 

\subsection{Mixed case}
\label{subsec-mixed}

Finally, let us consider the case where both types of coupling, kinetic and axial, have an important impact on the resulting gauge field. Naively, one would expect that this happens when $\kappa_2 \sim \kappa_1$. However, in this case we still find a strong domination of the kinetic coupling; this situation was discussed in the previous subsection. Therefore, we must consider $\kappa_2 > \kappa_1$, although by only 1 or 2 orders of magnitude. As previously, we can separate two stages: (i)~$\phi \gg M_{\mathrm{P}}\sqrt{3/2}\ln [1+(\kappa_1/\xi_s)^{-1/2}]$ when the second term in the kinetic coupling function~(\ref{I1}) dominates and (ii)~the opposite, $\phi \ll M_{\mathrm{P}}\sqrt{3/2}\ln [1+(\kappa_1/\xi_s)^{-1/2}]$, when $I_1\simeq 1$. The former regime shows the same features as in the kinetic-dominated case considered in Sec.~\ref{subsec-kinetic}, while the latter stage is very similar to the axial-dominated case discussed in Sec.~\ref{subsec-axial}. Taking into account the drawbacks of both previously considered cases, it is easy to conclude that we should take such values of $\kappa_1$ and $\kappa_2$ for which the transition between stages (i) and (ii) happens well before the end of inflation (to allow for a significant gauge-field amplification) and at the same time at less than 50--60 $e$-foldings from the end of inflation (so that backreaction does not spoil the perturbation modes relevant for CMB observations). 

For the Starobinsky potential (\ref{potential-general}), employing the slow-roll approximation and neglecting  backreaction from the produced gauge fields, we find the following dependence of the inflaton field on the number of $e$-foldings before the end of inflation (here $N=\ln\frac{a}{a_{\mathrm{end}}}<0$):
\begin{multline}
\frac{d\phi}{dN}=-M_{\mathrm{P}}^2 \frac{V'(\phi)}{V(\phi)}\approx -2\sqrt{\frac{2}{3}}M_{\mathrm{P}} e^{-\sqrt{\frac{2}{3}}\frac{\phi}{M_{\mathrm{P}}}}
\\ \Rightarrow \qquad e^{\sqrt{\frac{2}{3}}\frac{\phi}{M_{\mathrm{P}}}}\simeq \frac{4}{3}|N|.
\end{multline}
Using this result, we can  estimate the range of the coupling parameter $\kappa_1$ for which the transition occurs at $0\ll |N|\ll 50$. Taking the relevant range to be $15\lesssim |N|\lesssim 35$, we get the following result:
\begin{equation}
    \frac{\kappa_1}{\xi_s} \simeq \frac{9}{16 N^2}\sim (0.5 - 2.5)\times 10^{-3}.
\end{equation}
At the same time, $\kappa_2$ must be much greater than $\kappa_1$ in order to achieve the strong gauge-field amplification in the second stage. For definiteness, in our numerical computations we take $\kappa_1/\xi_s=1.5\times 10^{-3}$ which lies in the desired range and $\kappa_2/\xi_s=0.4$. For these values of coupling parameters, it suffices to truncate the system of equations of the gradient-expansion formalism at the order $n_{\mathrm{max}}=150$.

\begin{figure*}[ht]
\centering
\includegraphics[width=0.46\linewidth]{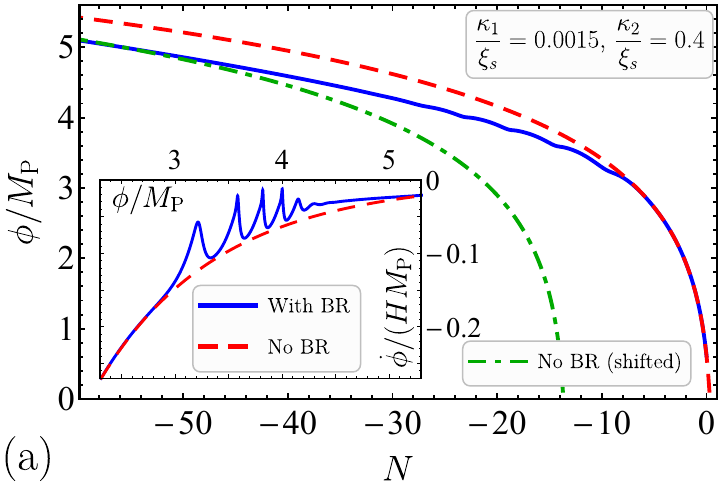}
\hspace{3mm}
\includegraphics[width=0.49\linewidth]{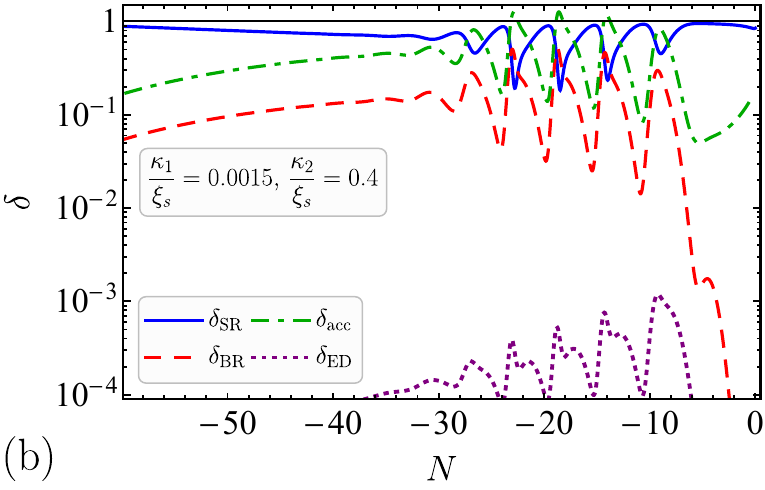}
\caption{(a) The dependence of the inflaton field $\phi$ on the number of $e$-foldings counted from the end of inflation in the absence of gauge fields (red dashed line) and for the case $\kappa_{1}/\xi_s=1.5\times 10^{-3}$, $\kappa_{2}/\xi_{s}=0.4$ where the backreaction occurs (blue solid line). The green dashed-dotted line shows the inflaton evolution without backreaction shifted by approximately 14 $e$-foldings so that it matches the blue curve far from the end of inflation. The inset shows a part of the phase portrait of the inflaton field in dimensionless coordinates [$\phi/M_{\mathrm{Pl}},\,\dot{\phi}/(HM_{\mathrm{Pl}})$]. (b) The $\delta$-parameters defined in Eq.~(\ref{delta-ED}), (\ref{delta-acc})--(\ref{delta-BR}) as functions of the number of $e$-foldings: the slow-roll parameter $\delta_{\mathrm{SR}}$ is shown by the blue solid line, the backreaction parameter $\delta_{\mathrm{BR}}$\,---\,by the red dashed line, the inflaton acceleration parameter $\delta_{\mathrm{acc}}$\,---\,by the green dashed-dotted line, and the energy-density parameter $\delta_{\mathrm{ED}}$\,---\,by the purple dotted line.}
\label{fig-inflaton-mix}
\end{figure*}

Figure~\ref{fig-inflaton-mix}(a) shows the evolution of the inflaton field in the presence of gauge fields (blue solid line) compared to the case without gauge fields (red dashed and green dashed-dotted lines). Here we observe that the inflaton field tends to match the free solution at more than 40 $e$-foldings from the end of inflation and during the last 10 $e$-foldings of inflation. This is also clearly seen from Fig.~\ref{fig-inflaton-mix}(b), where the parameter $\delta_{\mathrm{SR}}$ (shown by the blue solid line) tends to unity in these two limiting cases thus manifesting the slow-roll regime. In the intermediate stage, however, backreaction strongly affects the inflaton evolution, see, e.g., the strong modification of the inflaton phase portrait shown in the inset in Fig.~\ref{fig-inflaton-mix}(a) or the red dashed curve for the parameter $\delta_{\mathrm{BR}}$ in Fig.~\ref{fig-inflaton-mix}(b).  

\begin{figure*}[ht]
\centering
\includegraphics[width=0.99\linewidth]{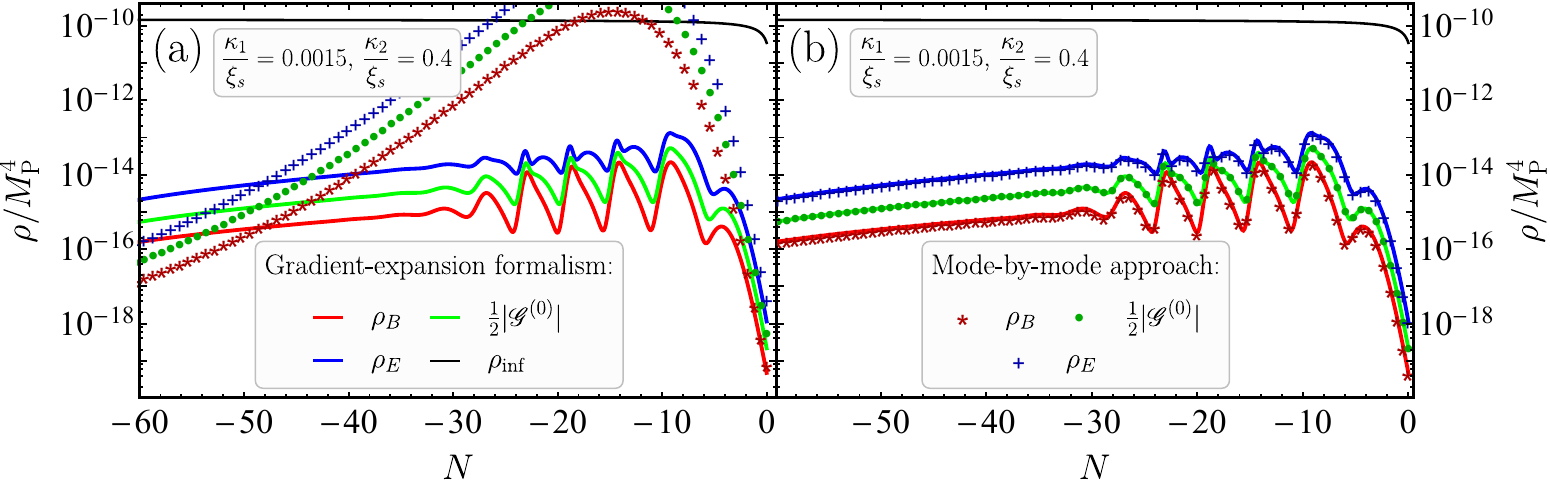}
\caption{The dependence of the energy densities on the number of $e$-foldings counted from the end of inflation in the case $\kappa_{1}/\xi_s=1.5\times 10^{-3}$, $\kappa_{2}/\xi_{s}=0.4$: magnetic energy density (red line), electric energy density (blue line), Chern-Pontryagin density (green line), and the energy density of the inflaton (black line). Dots marked by symbols show the corresponding quantities computed in the mode-by-mode approach: (a) when the backreaction is not taken into account, (b) when the backreaction modifies the inflaton evolution (based on the inflaton time dependence taken from the result of the gradient-expansion formalism).}
\label{fig-rho-mix}
\end{figure*}

The evolution of electric and magnetic energy densities as well as the quantity $\mathscr{G}^{(0)}$ (often called the Chern-Pontryagin density) of the generated gauge field are shown in Fig.~\ref{fig-rho-mix} by the solid lines. As in the axial-dominated case, here we also present the results obtained by considering separate Fourier modes of the gauge field in momentum space. The dotted lines in Fig.~\ref{fig-rho-mix}(a) correspond to the gauge field generated on the free inflaton background without backreaction while the similar lines in Fig.~\ref{fig-rho-mix}(b) are computed on the inflaton background affected by the backreaction (the time dependence of the inflaton and scale factor are obtained with the gradient-expansion formalism). Thus, Fig.~\ref{fig-rho-mix}(b) compares the values of the gauge field directly computed by the gradient-expansion formalism with the corresponding values obtained in the mode-by-mode treatment on the inflationary background taken from the gradient-expansion formalism. Although the latter value cannot be considered as a true reference solution since it is based on the partial results of the gradient-expansion formalism, nevertheless, it can be used to perform a consistency check of our method. As we see from Fig.~\ref{fig-rho-mix}(b), the dots match the solid lines with a good accuracy which numerically appears to be less than 1\,\% during the whole time interval under consideration (for more details, see Appendix~\ref{app-B}).

\begin{figure}[ht!]
\centering
\includegraphics[width=0.99\linewidth]{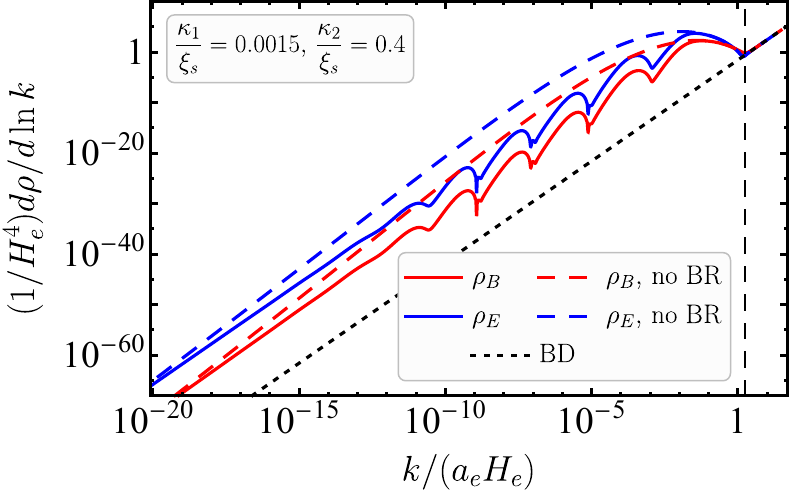}
\caption{The spectra of generated magnetic (red) and electric (blue) energy densities at the end of inflation in the case $\kappa_{1}/\xi_s=1.5\times 10^{-3}$, $\kappa_{2}/\xi_{s}=0.4$. The solid lines show the spectra in the case when the backreaction is taken into account while the dashed lines correspond to the gauge field generated on the free inflaton background.}
\label{fig-spectrum-mix}
\end{figure}

Finally, let us say a few words about the spectra of generated fields. The magnetic and electric spectral energy densities are shown by the blue and red lines in Fig.~\ref{fig-spectrum-mix}. The dashed lines show the spectra generated on the free inflaton background while the solid lines show the spectra in the case when the backreaction is taken into account. In full accordance with the inflaton evolution, one can distinguish three regions with different properties. For a better understanding, let us consider the behavior of the parameter $\xi$ introduced in Eq.~(\ref{xi-s}), which determines the gauge-field production if $\kappa_2\gg \kappa_1$. Using the slow-roll approximation for $\dot{\phi}$ and $H$, it can be expressed as
\begin{equation}
    \xi(\phi)=\frac{I_{2}^{\prime}(\phi)\dot{\phi}}{2H I_1}\simeq-\frac{4}{3}\frac{\kappa_2}{\xi_s} e^{\sqrt{\frac{2}{3}}\frac{\phi}{M_{\mathrm{P}}}}\bigg[1+\frac{\kappa_1}{\xi_s}\Big(e^{\sqrt{\frac{2}{3}}\frac{\phi}{M_{\mathrm{P}}}}-1\Big)^2\bigg]^{-1}.
\end{equation}

For modes crossing the horizon at more than 30 $e$-foldings before the end of inflation (when the inflaton field decreases monotonically) the spectral curves are monotonic and approach the unperturbed spectra in the limit of long-wavelength modes $k\to 0$. In this regime, the second term in the coupling function $I_1$ dominates; therefore, the parameter $\xi$ takes the form:
\begin{equation}
    |\xi|\approx \frac{\kappa_2}{\kappa_1}\frac{1}{3\sinh^2[\phi/(\sqrt{6}M_{\mathrm{P}})]},
\end{equation}
which is an increasing function of time (since $\phi$ always decreases). As a result, the spectrum is blue-tilted with the spectral tilt close to $n_B=4$.

For modes which cross the horizon between 30 and 10 $e$-foldings prior to the end of inflation, when strong backreaction occurs, the spectrum shows an oscillatory pattern which corresponds to the inflaton oscillations in the backreaction regime. On average, the spectrum is also blue-tilted with $n_B\approx 4$.

Finally, the modes which cross the horizon during the second slow-roll phase (during the last 10 $e$-foldings of inflation) have a red-tilted spectrum. This follows from the fact that the kinetic coupling function $I_1 \simeq 1$ at this stage and the parameter $\xi$ has the form
\begin{equation}
    |\xi|\simeq \frac{4}{3}\frac{\kappa_2}{\xi_s} e^{\sqrt{\frac{2}{3}}\frac{\phi}{M_{\mathrm{P}}}},
\end{equation}
which is a decreasing function of time. Therefore, the earlier mode crosses the horizon, the stronger it is amplified. Although, the red-tilted spectrum leads to a larger value of the coherence length of the produced gauge field, the range of such modes is very limited and spans over 2--3 orders of magnitude at maximum. Therefore, one may expect the coherence length just 2--3 orders of magnitude larger than the horizon size at the end of inflation.

\subsection{Oscillatory behavior in the backreaction regime}
\label{subsec-oscillations}

\begin{figure}[ht!]
	\centering
	\includegraphics[width=0.99\linewidth]{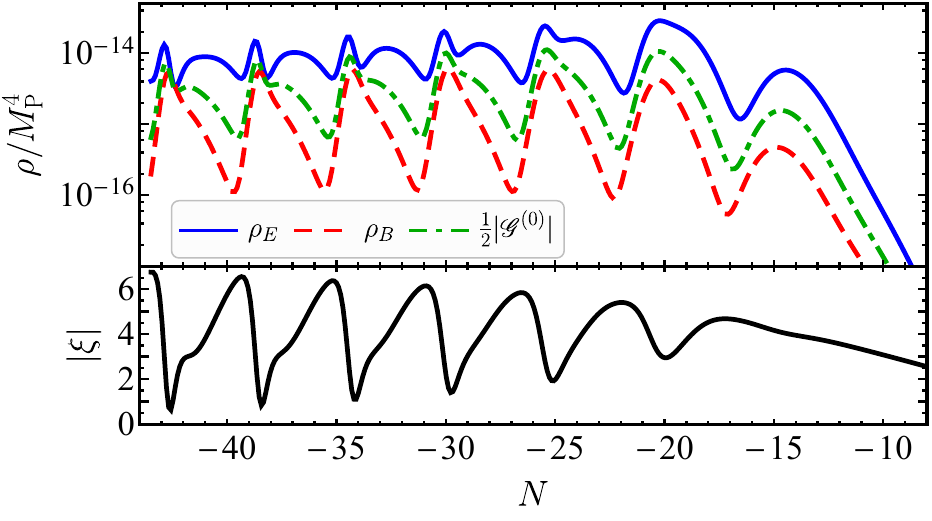}
	\caption{The dependence of energy densities (upper panel) and the gauge-field production parameter $|\xi|$ (lower panel) on the number of $e$-foldings from the end of inflation for $\kappa_{1}/\xi_s=10^{-4}$ and $\kappa_{2}/\xi_{s}=0.15$. A clear anticorrelation is observed (i.e., the oscillations of the two are shifted by a phase $\Delta\varphi\simeq \pi$). A period of oscillations in $e$-foldings, $T_N\simeq 4$ is roughly in accordance with the prediction $T_{N}\simeq 2\ln(\xi^{2}/2)\approx 4.15$ (for $|\xi|\simeq 4$) of Ref.~\cite{Domcke:2020}. \label{fig-rho-xi}}
\end{figure}

In this subsection we comment on oscillations of the inflaton and generated gauge fields in the backreaction regime. In Fig.~\ref{fig-rho-xi} we show the oscillations in the gauge-field energy density (upper panel) and in the parameter $|\xi|$ (lower panel) which occur in the axial-dominated case with $\kappa_1/\xi_s=10^{-4}$ and $\kappa_2/\xi_s=0.15$.

Note, that similar oscillations have already been reported in Ref.~\cite{Domcke:2020}, where it was shown that axial coupling of the inflaton field to gauge fields through the Chern-Simons term leads to a resonant enhancement of the gauge field production, resulting in oscillatory features in the inflaton velocity as well as in the gauge field spectrum. 
Since, the backreaction is relevant only in the axial-dominated case $\kappa_2\gg \kappa_1$, the similar arguments should be applicable  also in our model.

Remember that the backreaction first becomes important due to the term $(I_2^\prime /I_1)\mathscr{G}^{(0)}$ in the Klein-Gordon equation~(\ref{KGF-3}) which creates additional friction and slows down the inflaton motion. 
Naively, one would expect that this immediately suppresses the gauge-field production and the system finds another stable state (different from the slow-roll attractor) which smoothly evolves in time. However, the total generated gauge field value is retarded with respect to the changes in the inflaton. Indeed, the mode function with a given momentum $k$ gets enhanced only in a narrow time interval around its horizon crossing $k\sim k_{\mathrm{h}}$. For an axial-dominated coupling, this happens when $k\lesssim k_{\mathrm{h}}\simeq 2|\xi|aH$ and the maximal growth rate is achieved at the moment of time $t_k$ for which $k\simeq (1/2)k_{\mathrm{h}}(t_k)\simeq|\xi(t_k)|a(t_k)H(t_k)$. The growth rate and final amplitude of the mode $k$ is thus determined by the inflaton velocity (or the parameter $\xi$) at the moment of time $t_k$. However, the \textit{total} generated gauge field is a superposition of all modes with $k<2|\xi| aH$ and the leading contribution to it is made by modes which were enhanced at a slightly earlier time. In fact, in Ref.~\cite{Domcke:2020} it is shown that the leading contribution to the integral in $\mathscr{G}^{(0)}$ is made by modes around $k_{\mathrm{lead}}\simeq (2/|\xi|)aH$. These modes were enhanced $\Delta N\simeq \ln[k_{\mathrm{h}}/(2k_{\mathrm{lead}})]\simeq \ln(\xi^2/2)$ $e$-foldings before the actual moment of time. This causes the retardation in response of the gauge field to the variations of the inflaton velocity and leads to the oscillatory behavior. One may use this result to estimate the period of oscillations. It is easy to understand that the maximal amplitude would occur if the cause (the change in the inflaton velocity or in $\xi$ parameter) and a consequence (the corresponding change in the generated field) are opposite, i.e., they delay by $\pi$ in phase with respect to each other. Then, the period of oscillations in the $e$-foldings scale is just twice the retardation time
\begin{equation}
\label{period}
    T_{N}=2\Delta N\simeq 2\ln (\xi^2/2).
\end{equation}
These features can indeed be seen if we simultaneously plot the $\xi$ parameter and the generated fields, see Fig.~\ref{fig-rho-xi}. Their variations have opposite phase and the period of oscillations is in a good accordance with an estimate (\ref{period}).

We also note that in a recent paper \cite{Peloso:2022}, another attempt to analyze this behavior was made. The authors study the deviations from the ``would-be'' equilibrium solution in the absence of retardation and conclude that this solution is unstable. Moreover, the corresponding exponent is complex which corresponds to oscillations with an increasing amplitude around the equilibrium solution.

\section{Conclusion}
 \label{concl}
  
In this work, we study general features of inflationary magnetogenesis with strong backreaction from the produced gauge fields.
For this purpose, the gradient-expansion formalism previously proposed for the description of inflationary magnetogenesis in purely kinetic or purely axial coupling models, was extended to the case when both types of couplings are present. This formalism operates with a set of scalar bilinear functions (\ref{EE})--(\ref{BB}) which are quantum expectation values of scalar products of electric and/or magnetic field three-vectors with an arbitrary number of spatial derivatives. Since these quantities are defined in position space, they include all physically relevant Fourier modes of the gauge field at once and, thus, allow  to self-consistently take into account backreaction of the generated fields on the background evolution. The latter phenomenon makes the gauge-filed dynamics strongly nonlinear and couples all Fourier modes to each other; consequently, the standard mode-by-mode approach operating in momentum space becomes extremely complicated as it requires to evolve simultaneously a huge number of coupled Fourier modes.

In general, the vacuum expectation value involves contributions from all Fourier modes with momenta from zero to infinity. However, modes with large momenta (subhorizon modes) are almost unaffected by the coupling to the inflaton and quickly oscillate in time\,---\,they correspond to vacuum fluctuations of the gauge field. Thus, to capture the effect of gauge-field production due to kinetic and axial coupling it is important to separate it from the contributions of vacuum fluctuations. This was performed by taking into account only modes with momenta $k\leq k_{\mathrm{h}}$, where the threshold value $k_{\mathrm{h}}$ is defined in Eq.~(\ref{k-h-mixed}). This threshold is chosen such that a term in the mode equation~(\ref{eq-mode-z-2}) describing the free evolution of the mode is equal in absolute value to terms originating from the kinetic and axial couplings. The introduction of this momentum cutoff allowed us to extract from an infinite vacuum expectation value a finite part containing information about the produced gauge field.

Since the threshold momentum is an increasing function of time during inflation, the equations of motion for the quantities $\mathscr{E}^{(n)}$, $\mathscr{G}^{(n)}$, $\mathscr{B}^{(n)}$ must take into account the additional time variation due to new modes which cross the horizon and start contributing to these quantities. This is done by introducing boundary terms on the right-hand side of Eqs.~(\ref{eq-EE})--(\ref{eq-BB}). Analyzing the evolution of the mode function around the moment of horizon crossing, we derived explicit expressions for the boundary terms in terms of  Whittaker functions.

Using this newly introduced gradient-expansion formalism we studied the gauge-field production in extended Starobinsky model which includes the gauge field nonminimally coupled to gravity. This model is constructed in such a way that, rewritten in Einstein frame, it possesses a number of important features: (i) it acquires a scalar degree of freedom (similarly to the usual Starobinsky model) which plays the role of the inflaton and has an asymptotically flat potential as favored by CMB observations~\cite{Planck:2018-infl}; (ii) its Lagrangian is quadratic in the gauge fields and has the form of kinetic and axial couplings to the inflaton; (iii) for positive values of the coupling parameter $\kappa_1$ it avoids the strong coupling problem during inflation; (iv) the absence of higher order terms in the gauge fields allows us to study the gauge field nonperturbatively and take into account its backreaction on the evolution of the inflaton. Thus, this model is perfectly suitable for analysis by the gradient-expansion formalism.

In this model, significant gauge-field production  occurs only if the axial coupling is much stronger than the kinetic one; i.e., $\kappa_2 \gg \kappa_1$. However, this inequality does not imply that the kinetic coupling plays no role and can be neglected. In the mode equation~(\ref{eq-mode-z-2}), parameter $s=\frac{\dot{I}_{1}}{2H I_{1}}+\frac{\ddot{I_1}}{2H^2 I_1}-\frac{\dot{I}_{1}^2}{4H^2 I_{1}^2},$ is indeed much smaller than the  parameter $\xi=\frac{\dot{I}_{2}}{2H I_1}$. Nevertheless, the kinetic coupling function $I_1$ enters the expression for $\xi$ in the denominator. Therefore, it modulates the magnitude of the axial coupling and makes an impact on the resulting gauge field. This property is extremely important as it helps to suppress  gauge-field production (and its backreaction on the background evolution) at 50--60 $e$-foldings before the end of inflationin order not to spoil the behavior of perturbation modes relevant for the CMB.

Since the coupling functions $I_{1,2}$ given by Eq.~(\ref{I1}) are decreasing in time, the generated field is also decreasing toward the end of inflation. Therefore, if the produced gauge field at the end of inflation is not extremely small, backreaction is typically occurring when we go back into the past. We would like to point out a few general features of the backreaction regime which were observed. First, backreaction always occurs due to terms on the right-hand side of the Klein-Gordon equation~(\ref{KGF-3}) for the inflaton field. Moreover, contrary to a rather common lore in the literature, it is completely irrelevant in the Friedmann equation, because the energy density of the produced gauge field remains suppressed by 3--4 orders compared to  the energy density of inflaton. Therefore, the backreaction criterion based on the gauge-field contribution to the total energy density is incorrect.
Here we have shown that even when the gauge field energy density is much smaller than the one of the inflaton, backreaction is relevant. 

Second, the time behavior of the inflaton velocity and generated gauge-field energy density in the backreaction regime shows regular oscillations in their absolute values (although the sign of $\dot{\phi}$ does not change in course of these oscillations meaning that the inflaton remains a monotonically decreasing function of time). These oscillations originate from the retardation between the changes in the inflaton field and the corresponding response in the gauge field. This feature was known in the literature before~\cite{Domcke:2020,Peloso:2022}, however, to the best of our knowledge, such a regular oscillatory pattern in a realistic inflationary model is reported for the first time. 

Third, despite the oscillatory behavior in backreaction regime, for any given value of the inflaton field its velocity is always smaller in absolute value than the corresponding velocity in the absence of backreaction. Therefore, it takes more time for the inflaton field to roll down to the minimum of its potential and finish the inflation stage. As a result, rather than spoiling it, backreaction {\em increases} the duration of inflation by several $e$-foldings, depending on the model parameters. This behavior, in some sense similar to the ultra-slow-roll regime in potentials with an inflection point, may be important for generation of strong small-scale scalar perturbations and for primordial black hole production.

Forth, backreaction has a strong impact on the spectra of the generated gauge fields. As it was shown in Ref.~\cite{Durrer:2022emo}, in the absence of backreaction, by choosing a sufficiently large value of parameter $\kappa_2$, it is possible to achieve a scale-invariant or red-tilted magnetic power spectrum and, thus, obtain a large correlation scale for  the produced gauge field. Backreaction drastically changes this behavior. For gauge-field modes which cross the horizon in the backreaction regime, the spectrum also reveals oscillatory behavior and the average spectral index appears to be close to $n_B=4$. Since  backreaction turns off a few $e$-foldings before the end of inflation, one can still get a part of the spectrum which is red-tilted, although for a limited range of modes spanning over 2--3 orders of magnitude. Therefore, the resulting coherence length of the produced gauge fields may be at most 2--3 orders of magnitude larger than the horizon size at the end of inflation.

All particular features mentioned above lead us to the conclusion that theories with decreasing coupling functions, although they seemingly imply a potentially larger coherence length of  the generated gauge fields, unavoidably run into a backreaction regime which has two major consequences: (i) It limits the resulting magnitude of the produced gauge field because in the backreaction regime, the gauge-field energy density is smaller than that of the inflaton by several orders of magnitude, while once backreaction switches off the field can only decrease. (ii) Backreaction does not allow for a significant increase in the magnetic correlation length as it turns a red-tilted spectrum into a blue-tilted one. 
In our previous work~\cite{Durrer:2022emo}, where the same type of coupling functions arose in the Higgs-Starobinsky inflationary model, but the self-consistent description of the backreaction was not possible (because of higher powers of gauge fields present in the action in that model), we have concluded that it is possible to obtain a scale-invariant or even red-tilted spectrum  that implies very large correlation length of the generated fields. A nonperturbative analysis of the backreaction performed in the present work, unfortunately, made the constraints on the magnetic correlation length much more severe than in Ref.~\cite{Durrer:2022emo}. 

Let us finally emphasize that the gradient-expansion formalism developed in this work is not restricted to this type of models with decreasing coupling functions. It can be used for the description of gauge-field production in a wide range of models of inflationary magnetogenesis described by the action~(\ref{action-kinetic-axial}). Since the generated fields are extremely strong during inflation, it would be interesting to take into account also the Schwinger pair production. This may be interesting also for the model considered in the present study because the Schwinger effect in some cases helps to avoid backreaction~\cite{Gorbar:2021rlt}. Another important problem is to study the impact of produced gauge fields on the spectra of primordial scalar and tensor perturbations. Requiring these spectra to be in accordance with CMB and BBN observations, one can strongly constrain the parameter space of magnetogenesis models (see, e.g., Refs.~\cite{Ferreira:2014,Barnaby:2012,Adshead:2018doq,Adshead:2019igv,Adshead:2019lbr}). We plan to address these issues elsewhere.

\begin{acknowledgments}
We would like to thank A.V.~Lysenko for her help in performing numerical computations and useful discussions.
The work of O.S. was supported by the National Research Foundation of Ukraine Project No.~2020.02/0062.
The work of R.D. and S.V. is supported by Swiss National Science Foundation Grant No.~SCOPE IZSEZ0 206908. The work of S.V. is supported by Scholars at Risk  and Swiss National Science Foundation. 
O.S. is grateful to Prof. Kai Schmitz and to all members of the Particle Cosmology group for their kind hospitality in University of M\"{u}nster where the final part of this work was done. The work of O.S. was sustained by a Philipp-Schwartz fellowship of the University of M\"{u}nster.
\end{acknowledgments}

\appendix
\section{PROPERTIES OF THE WHITTAKER FUNCTIONS}
\label{app-Whittaker}

The differential equation
\begin{equation}
\label{Whittaker-eq}
    \frac{d^{2}w}{dy^{2}}+\left(-\frac{1}{4}+\frac{\varkappa}{y}+\frac{1/4-\mu^{2}}{y^{2}}\right)w=0
\end{equation}
is known as the Whittaker equation and has two linearly independent solutions $M_{\kappa,\mu}$ and $W_{\kappa,\mu}$. 
For the our purposes, however, only the function $W_{\kappa,\mu}$ is relevant. It can be expressed in terms of the Tricomi confluent hypergeometric function $U$ as follows:
\begin{equation}
	W_{\varkappa,\mu}(y)=e^{-y/2}y^{\mu+1/2}U(\mu-\varkappa+1/2;1+2\mu;y).
\end{equation}

Using Eq.~(13.5.2) in Ref.~\cite{AbramowitzStegun}, one can derive the following asymptotical expression of the Whittaker function at $|y|\to \infty$:
\begin{equation}
\label{W-asym}
    W_{\varkappa,\mu}(y)=e^{-y/2} y^{\varkappa} [1+O(y^{-1})].
\end{equation}

The mode equation (\ref{eq-mode-z-2}) has the form of the Whittaker equation (\ref{Whittaker-eq}) 
with $\kappa=-i\lambda \xi$, $\mu=\pm\sqrt{1/4+s}$, and $y=2iz$. Its solution must satisfy the Bunch-Davies vacuum boundary condition 
(\ref{BD}). Comparing it with Eq.~(\ref{W-asym}), we conclude that the $W$ function indeed has the correct asymptotic. Therefore, the solution to the mode equation has the form (\ref{A_Whittaker}).

In the derivation of boundary terms we used the expression for the derivative of the Whittaker $W$ function which is given by (see Eq.~(13.4.33) in Ref.~\cite{AbramowitzStegun})
\begin{equation}
\label{Whittaker-derivative}
	y\frac{d}{dy}W_{\varkappa,\mu}(y)=(y/2-\kappa)W_{\varkappa,\mu}(y)-W_{\varkappa+1,\mu}(y).
\end{equation}

\section{DETAILS OF NUMERICAL COMPUTATIONS}
\label{app-B}

In this Appendix, we give more details on the numerical analysis of the model discussed in the main text. In particular, we discuss two approximations which can be used in order to simplify the numerical implementation of the boundary terms.

Equations~(\ref{E_p_d})--(\ref{B-lambda}) give full expressions for the boundary terms. There are two ingredients in these expressions which may strongly complicate the numerical analysis: the presence of Whittaker functions and the factor $d\ln k_{\mathrm{h}}/dt$. Let us consider them separately.

\begin{figure*}[ht]
	\centering
	\includegraphics[width=0.48\linewidth]{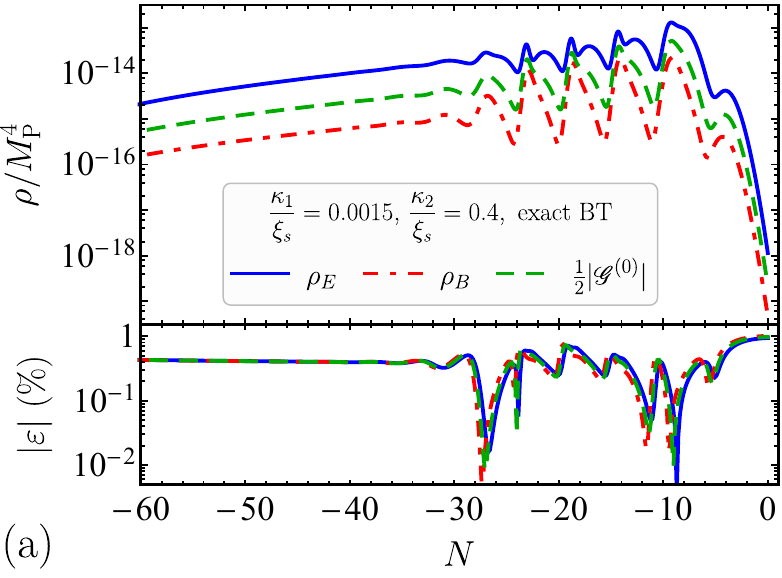}
	\hspace{3mm}
	\includegraphics[width=0.48\linewidth]{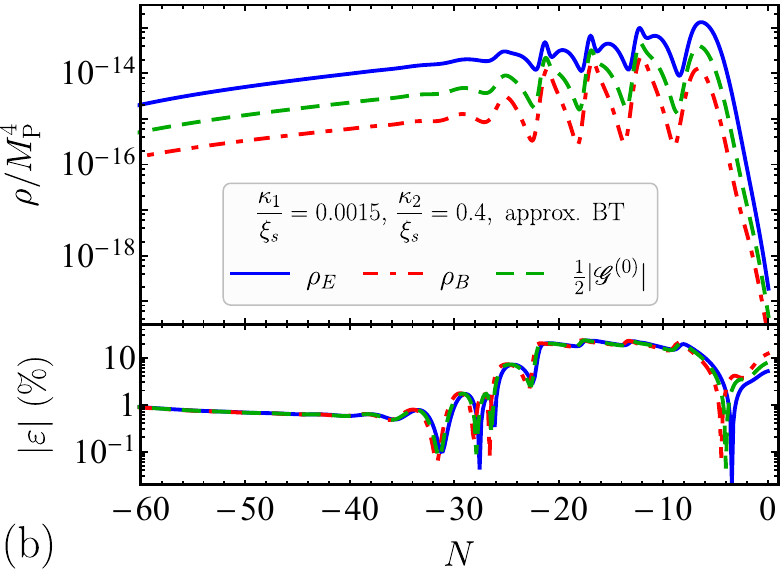}
	\caption{Upper panels: The dependence of the energy densities on the number of $e$-foldings counted from the end of inflation in the case $\kappa_{1}/\xi_s=1.5\times 10^{-3}$, $\kappa_{2}/\xi_{s}=0.4$: magnetic energy density (red line), electric energy density (blue line), and the Chern-Pontryagin density (green line). Lower panels show the relative discrepancy between the result obtained using the gradient expansion formalism and the corresponding mode-by-mode solution (based on the inflaton time dependence taken from the result of the gradient-expansion formalism). Panel~(a) corresponds to the case where $dk_\mathrm{h}/dt$ is computed as in Eq.~(\ref{dkh-dt}) while panel~(b) shows the result when an approximation (\ref{dkh-dt-approx}) was used.}
	\label{fig-accuracy-comparison}
\end{figure*}

The Whittaker $W$ function belongs to the class of confluent hypergeometric functions (see Appendix~\ref{app-Whittaker}), i.e., is a special function. Therefore, its presence in the numerical program requires using some additional packages which include this function (e.g., in C++ or Python) and/or in some cases requires computations with a precision exceeding the standard machine precision (e.g., this happens in Wolfram \textit{Mathematica} for large values of parameter $\xi$; see Ref.~\cite{Gorbar:2021rlt}). In such a case, it could be useful to have approximate expressions for the boundary terms only involving elementary functions. For the case of a pure axial coupling such expressions have been derived in Ref.~\cite{Gorbar:2021rlt} both in the absence and presence of the Schwinger effect. In the present work, expressions in Eqs.~(\ref{E-lambda})--(\ref{B-lambda}) are more complicated than the pure axial coupling case; however, if we assume that $\dot{I}_1/I_1$ is a slowly varying function and neglect its time derivative, the expression for the horizon-crossing momentum in Eq.~(\ref{k-h-fin}) takes the form
\begin{multline}
\label{k-h-approx}
k_{\mathrm{h}}(t)\approx\max\limits_{t'\leq t}\Big\{a(t')H(t')\Big[|\xi(t')|\\
+\sqrt{\xi^2(t')+|\sigma^2(t')+\sigma(t')|} \Big] \Big\},
\end{multline}
where $\sigma(t)$ is the parameter introduced in Eq.~(\ref{sigma-par}). This expression now coincides (up to an additional modulus sign) with the corresponding expression for the horizon-crossing momentum in the case of purely axial coupling in the presence of Schwinger effect; see Eq.~(51) in Ref.~\cite{Gorbar:2021rlt}. The correspondence between these two cases can be revealed replacing the Schwinger conductivity by the quantity $\dot{I}_1/I_1$. The only difference between these cases is that the conductivity is always positive while $\dot{I}_1/I_1$ is typically negative when  gauge-field amplification occurs [this explains the presence of additional modulus sign under the square root in Eq.~(\ref{k-h-approx})]. The above mentioned correspondence allowed us to use in the present work approximate expressions for the functions $E_\lambda$, $G_\lambda$, $B_\lambda$ derived in Ref.~\cite{Gorbar:2021rlt}, namely Eqs.~(B15)--(B20), with a replacement $s^2+s \to |\sigma^2+\sigma|$. These expressions have an error less than 0.5\,\% (compared to corresponding expressions in terms of Whittaker functions) for $|\xi|>4$. For lower values of $|\xi|$ we used the boundary terms expressed through the Whittaker functions.

Another issue that may cause numerical complications is the derivative $d\ln k_\mathrm{h}/dt$. Indeed, the momentum of horizon-crossing mode, $k_\mathrm{h}(t)$, given by Eq.~(\ref{k-h-fin}) is a maximal value of a certain expression over the whole time interval from $t'=0$ until time $t$. A possible way to compute the time derivative of such an expression is the following~\cite{Sobol:2020lec}. Let us denote
\begin{equation}
    k_\mathrm{h}=\max\limits_{t'\leq t} [y(t')].
\end{equation}
Then,
\begin{equation}
\label{dkh-dt}
    \frac{dk_\mathrm{h}}{dt}=\frac{dy(t)}{dt}\cdot \theta\Big[\frac{dy(t)}{dt}\Big]\cdot \theta[y(t)-k_\mathrm{h}(t)],
\end{equation}
where $\theta[x]$ is the Heaviside unit step function. Since the function $y(t)$ is already rather complicated, Eq.~(\ref{dkh-dt}) appears to be very cumbersome and, since it appears in all equations for the bilinear functions, it requires a lot of computational resources. In axial-dominated and mixed cases, we used an approximation $y(t)=2aH|\xi|$ which is well-justified since $|\sigma|\ll |\xi|$ in these cases. The consistency check performed by comparison with the result of the mode-by-mode approach shows that the relative discrepancy is typically less than 1\,\%; see Fig.~\ref{fig-accuracy-comparison}(a).

Moreover, one can simplify $dk_\mathrm{h}/dt$ even further and take into account only the time derivative of the scale factor $a$ because it has the fastest (quasiexponential) change in time. In this approximation,
\begin{equation}
    \label{dkh-dt-approx}
    \frac{dk_\mathrm{h}}{dt}\approx H k_\mathrm{h}.
\end{equation}
This approximation ignores the fact that in the backreaction regime $\xi$ may also change very fast and the function $y(t)$ may be even nonmonotonic. However, it allows to accelerate computations by several times. The consistency check with the mode-by-mode solution shows a discrepancy of up to 20\,\%--25\,\%. Such an accuracy should be enough for a qualitative analysis and order-of-magnitude estimates; however, in order to obtain a more precise result one must use Eq.~(\ref{dkh-dt}). 

We compare the consistency of numerical results obtained by the gradient-expansion formalism in the mixed case $\kappa_{1}/\xi_s=1.5\times 10^{-3}$, $\kappa_{2}/\xi_{s}=0.4$ in Fig.~\ref{fig-accuracy-comparison}, where panel~(a) corresponds to the time derivative $dk_\mathrm{h}/dt$ computed according to Eq.~(\ref{dkh-dt}) and panel~(b) shows the result of approximation~(\ref{dkh-dt-approx}).

Finally, let us mention that in some cases the program which solves the system of equations of the gradient-expansion formalism fails to proceed toward the end of inflation because the number of equations is very large (it reaches 500 in the case of strong backreaction) and the computational error accumulates rapidly. In this case, the following algorithm has to be used  to reach the end of inflation:
\begin{enumerate}
    \item Using the gradient-expansion formalism with a certain truncation order $n_\mathrm{max,1}$ (chosen in such a way that the result does not change when $n_\mathrm{max,1}$ is increased) obtain the numerical solution on the time interval from $t=0$ to a certain $t=t_1<t_e$ which is still well before the time when the program breaks down.
    \item Generate a table of values of all bilinear functions and background quantities at $t=t_1$ and use it as an initial condition for a new run of the gradient-expansion formalism with lower truncation order $n_\mathrm{max,2}<n_\mathrm{max,1}$.
    \item Repeat this procedure a few times if needed. The full solution is then obtained as a piecewise function. Its consistency with the mode-by-mode solution has to be checked in the same way as for a single continuous function.
\end{enumerate}
This procedure was used in our computations in the axial-dominated and mixed cases where the derivative $dk_\mathrm{h}/dt$ was calculated according to Eq.~(\ref{dkh-dt}). For the simplified case employing Eq.~(\ref{dkh-dt-approx}) such a problem never occurred and it was possible to reach the end of inflation in a single run of the code.


\begin{thebibliography}{99}


\bibitem{Starobinsky:1980} A.A.~Starobinsky,
A new type of isotropic cosmological models without singularity,
\href{https://doi.org/10.1016/0370-2693(80)90670-X}{Phys. Lett. \textbf{91B}, 99 (1980)}. 

\bibitem{Guth:1981} A.H.~Guth,
The inflationary Universe: A possible solution to the horizon and flatness problems,
\href{https://doi.org/10.1103/PhysRevD.23.347}{Phys. Rev. D \textbf{23}, 347 (1981)}.

\bibitem{Linde:1982} A.D.~Linde,
A new inflationary Universe scenario: A possible solution of the horizon, flatness, homogeneity, isotropy and primordial monopole problems,
\href{https://doi.org/10.1016/0370-2693(82)91219-9}{Phys. Lett. B \textbf{108}, 389 (1982)}.

\bibitem{Starobinsky:1982} A.A.~Starobinsky,
Dynamics of phase transition in the new inflationary Universe scenario and generation of perturbations,
\href{https://doi.org/10.1016/0370-2693(82)90541-X}{Phys. Lett. B \textbf{117}, 175 (1982)}.

\bibitem{Albrecht:1982} A.~Albrecht and P.J.~Steinhardt,
Cosmology for Grand Unified Theories with radiatively induced symmetry breaking,
\href{https://doi.org/10.1103/PhysRevLett.48.1220}{Phys. Rev. Lett. \textbf{48}, 1220 (1982)}. 

\bibitem{Linde:1983} A.D.~Linde,
Chaotic inflation,
\href{https://doi.org/10.1016/0370-2693(83)90837-7}{Phys. Lett. B \textbf{129}, 177 (1983)}.


\bibitem{Starobinsky:1979} A.A.~Starobinsky,
Spectrum of relict gravitational radiation and the early state of the universe,
JETP Lett. \textbf{30}, 682 (1979).

\bibitem{Mukhanov:1981} V.F.~Mukhanov and G.V.~Chibisov,
Quantum fluctuations and a nonsingular Universe,
JETP Lett. \textbf{33}, 532 (1981).

\bibitem{Mukhanov:1982} V.F.~Mukhanov and G.V.~Chibisov,
The Vacuum energy and large scale structure of the universe,
Sov. Phys. JETP \textbf{56}, 258 (1982).

\bibitem{Guth:1982} A.H.~Guth and S.Y.~Pi,
Fluctuations in the new inflationary Universe,
\href{https://doi.org/10.1103/PhysRevLett.49.1110}{Phys. Rev. Lett. \textbf{49}, 1110 (1982)}.

\bibitem{Hawking:1982} S.W.~Hawking,
The development of irregularities in a single bubble inflationary Universe,
\href{https://doi.org/10.1016/0370-2693(82)90373-2}{Phys. Lett. B \textbf{115}, 295 (1982)}.

\bibitem{Bardeen:1983} J.M.~Bardeen, P.J.~Steinhardt, and M.S.~Turner,
Spontaneous creation of almost scale-free density perturbations in an inflationary Universe,
\href{https://doi.org/10.1103/PhysRevD.28.679}{Phys. Rev. D \textbf{28}, 679 (1983)}.




\bibitem{Planck:2013-infl} P.A.R.~Ade \textit{et al.} (Planck Collaboration),
Planck 2013 results. XXII. Constraints on inflation,
\href{https://doi.org/10.1051/0004-6361/201321569}{Astron. Astrophys. \textbf{571}, A22 (2014)}
[\href{https://arxiv.org/abs/1303.5082}{arXiv:1303.5082 [astro-ph.CO]}].

\bibitem{Planck:2015-infl} P.A.R.~Ade \textit{et al.} (Planck Collaboration),
Planck 2015 results. XX. Constraints on inflation,
\href{https://doi.org/10.1051/0004-6361/201525898}{Astron. Astrophys. \textbf{594}, A20 (2016)}
[\href{https://arxiv.org/abs/1502.02114}{arXiv:1502.02114 [astro-ph.CO]}].

\bibitem{Planck:2018-infl} Y.~Akrami  \textit{et al.} (Planck Collaboration),
Planck 2018 results. X. Constraints on inflation, \href{https://doi.org/10.1051/0004-6361/201833887}{Astron. Astrophys. \textbf{641}, A10 (2020)}
[\href{https://arxiv.org/abs/1807.06211}{arXiv: 1807.06211 [astro-ph.CO]}].

\bibitem{Martin:2018} J.~Martin,
The Theory of Inflation,
\href{https://doi.org/10.3254/ENFI200008}{Proc. Int. Sch. Phys. Fermi \textbf{200}, 155 (2020)}
[\href{https://arxiv.org/abs/1807.11075}{arXiv:1807.11075 [astro-ph.CO]}].

\bibitem{Martin:2014} J.~Martin, C.~Ringeval and V.~Vennin,
Encyclop\ae{}dia Inflationaris,
\href{https://doi.org/10.1016/j.dark.2014.01.003}{Phys. Dark Univ. \textbf{5-6}, 75-235 (2014)}
[\href{https://arxiv.org/abs/1303.3787}{arXiv:1303.3787 [astro-ph.CO]}].



\bibitem{Tavecchio:2010} F.~Tavecchio, G.~Ghisellini, L.~Foschini, G.~Bonnoli, G.~Ghirlanda, and P.~Coppi,
The intergalactic magnetic field constrained by Fermi/Large Area Telescope observations of the TeV blazar 1ES 0229+200,
\href{https://doi.org/10.1111/j.1745-3933.2010.00884.x}{Mon. Not. R. Astron. Soc. \textbf{406}, L70 (2010)}
[\href{https://arxiv.org/abs/1004.1329}{arXiv: 1004.1329 [astro-ph.CO]}].
	
\bibitem{Ando:2010}	S.~Ando and A.~Kusenko,
Evidence for gamma-ray halos around active galactic nuclei and the first measurement of intergalactic magnetic fields,
\href{https://doi.org/10.1088/2041-8205/722/1/L39}{Astrophys. J. Lett. \textbf{722}, L39 (2010)}
[\href{https://arxiv.org/abs/1005.1924}{arXiv: 1005.1924 [astro-ph.HE]}].
	
\bibitem{Neronov:2010} A.~Neronov and I.~Vovk,
Evidence for strong extragalactic magnetic fields from Fermi observations of TeV blazars,
\href{https://doi.org/10.1126/science.1184192}{Science \textbf{328}, 73 (2010)}
[\href{https://arxiv.org/abs/1006.3504}{arXiv: 1006.3504 [astro-ph.HE]}].
	
\bibitem{Dolag:2010} K.~Dolag, M.~Kachelriess, S.~Ostapchenko, and R.~Tom\`{a}s,
Lower limit on the strength and filling factor of extragalactic magnetic fields,
\href{https://doi.org/10.1088/2041-8205/727/1/L4}{Astrophys. J. Lett. \textbf{727}, L4 (2011)}
[\href{https://arxiv.org/abs/1009.1782}{arXiv: 1009.1782 [astro-ph.HE]}].
	
\bibitem{Dermer:2011} C.D.~Dermer, M.~Cavadini, S.~Razzaque, J.D.~Finke, J.~Chiang, and B.~Lott,
Time delay of cascade radiation for TeV blazars and the measurement of the intergalactic magnetic field,
\href{https://doi.org/10.1088/2041-8205/733/2/L21}{Astrophys. J. Lett. \textbf{733}, L21 (2011)}
[\href{https://arxiv.org/abs/1011.6660}{arXiv: 1011.6660 [astro-ph.HE]}].
	
\bibitem{Taylor:2011} A.M.~Taylor, I.~Vovk, and A.~Neronov,
Extragalactic magnetic fields constraints from simultaneous GeV-TeV observations of blazars,
\href{https://doi.org/10.1051/0004-6361/201116441}{Astron. Astrophys. \textbf{529}, A144 (2011)}
[\href{https://arxiv.org/abs/1101.0932}{arXiv: 1101.0932 [astro-ph.HE]}].
	
\bibitem{Caprini:2015} C.~Caprini and S.~Gabici,
Gamma-ray observations of blazars and the intergalactic magnetic field spectrum,
\href{https://doi.org/10.1103/PhysRevD.91.123514}{Phys. Rev. D \textbf{91}, 123514 (2015)}
[\href{https://arxiv.org/abs/1504.00383}{arXiv: 1504.00383 [astro-ph.CO]}].

\bibitem{MAGIC:2022} V.A.~Acciari \textit{et al.} (MAGIC Collaboration),
A lower bound on intergalactic magnetic fields from time variability of 1ES 0229+200 from MAGIC and Fermi/LAT observations,
\href{https://doi.org/10.1051/0004-6361/202244126}{Astron. Astrophys. \textbf{670}, A145 (2023)}
[\href{https://arxiv.org/abs/2210.03321}{arXiv:2210.03321 [astro-ph.HE]}].

	

\bibitem{Planck:2015-pmf} P.A.R. Ade \textit{et al.} (Planck Collaboration),
Planck 2015 results. XIX. Constraints on primordial magnetic fields,
\href{https://doi.org/10.1051/0004-6361/201525821}{Astron. Astrophys. \textbf{594}, A19 (2016)}
[\href{https://arxiv.org/abs/1502.01594}{arXiv:1502.01594 [astro-ph.CO]}].
	
\bibitem{Sutton:2017} D.R.~Sutton, C.~Feng, and C.L.~Reichardt,
Current and	future constraints on primordial magnetic fields,
\href{https://doi.org/10.3847/1538-4357/aa85e2}{Astrophys. J. \textbf{846}, 164 (2017)}
[\href{https://arxiv.org/abs/1702.01871}{arXiv:1702.01871 [astro-ph.CO]}].

\bibitem{Giovannini:2018b} M.~Giovannini,
Probing large-scale magnetism with the Cosmic Microwave Background,
\href{https://doi.org/10.1088/1361-6382/aab17d}{Classical Quantum Gravity  \textbf{35}, 084003 (2018)}
[\href{https://arxiv.org/abs/1712.07598}{arXiv:1712.07598 [astro-ph.CO]}].
	
\bibitem{Paoletti:2018} D.~Paoletti, J.~Chluba, F.~Finelli, and J.A.~Rubi\~{n}o-Mart\'{i}n,
Improved CMB anisotropy constraints on primordial magnetic fields from the post-recombination ionization history,
\href{https://doi.org/10.1093/mnras/sty3521}{Mon. Not. R. Astron. Soc. \textbf{484}, 185 (2019)}
[\href{https://arxiv.org/abs/1806.06830}{arXiv:1806.06830 [astro-ph.CO].}]

\bibitem{Brandenburg:2020vwp} A.~Brandenburg, R.~Durrer, Y.~Huang, T.~Kahniashvili, S.~Mandal, and S.~Mukohyama,
Primordial magnetic helicity evolution with a homogeneous magnetic field from inflation,
\href{https://doi.org/10.1103/PhysRevD.102.023536}{Phys. Rev. D \textbf{102}, 023536 (2020)}
[\href{https://arxiv.org/abs/2005.06449}{arXiv:2005.06449 [astro-ph.CO]}].


	
\bibitem{Bray:2018} J.D.~Bray and A.M.M.~Scaife,
An upper limit on the strength of the extragalactic MF from ultra-high-energy cosmic-ray anisotropy,
\href{https://doi.org/10.3847/1538-4357/aac777}{Astrophys. J. \textbf{861}, 3 (2018)}
[\href{https://arxiv.org/abs/1805.07995}{arXiv:1805.07995 [astro-ph.CO]}].

\bibitem{Neronov:2021} A.~Neronov, D.~Semikoz, and O.~Kalashev,
Limit on intergalactic magnetic field from ultra-high-energy cosmic ray hotspot in Perseus-Pisces region,
\href{https://arxiv.org/abs/2112.08202}{2112.08202 [astro-ph.HE]}.





\bibitem{Durrer:2013} R.~Durrer and A.~Neronov,
Cosmological magnetic fields: Their generation, evolution and observation, \href{https://doi.org/10.1007/s00159-013-0062-7}{Astron. Astrophys. Rev. \textbf{21}, 62 (2013)}
[\href{https://arxiv.org/abs/1303.7121}{arXiv: 1303.7121 [astro-ph.CO]}].
		
\bibitem{Parker:1968} L.~Parker,
Particle creation in expanding universes, \href{https://doi.org/10.1103/PhysRevLett.21.562}{Phys. Rev. Lett. \textbf{21}, 562 (1968)}.




\bibitem{Turner:1988} M.S.~Turner and L.M.~Widrow,
Inflation-produced, large-scale magnetic fields,
\href{https://doi.org/10.1103/PhysRevD.37.2743}{Phys. Rev. D \textbf{37}, 2743 (1988)}.
	
\bibitem{Ratra:1992} B.~Ratra,
Cosmological `seed' magnetic field from inflation,
\href{https://doi.org/10.1086/186384}{Astrophys. J. \textbf{391}, L1 (1992)}.
	
\bibitem{Garretson:1992} W.D.~Garretson, G.B.~Field, and S.M.~Carroll,
Primordial magnetic fields from pseudo-Goldstone bosons,
\href{https://doi.org/10.1103/PhysRevD.46.5346}{Phys. Rev. D {\bf 46}, 5346 (1992)}
[\href{https://arxiv.org/abs/hep-ph/9209238}{arXiv: hep-ph/9209238}].
	
\bibitem{Dolgov:1993} A.D.~Dolgov,
Breaking of conformal invariance and electromagnetic field generation in the Universe,
\href{https://doi.org/10.1103/PhysRevD.48.2499}{Phys. Rev. D \textbf{48}, 2499 (1993)}
[\href{https://arxiv.org/abs/hep-ph/9301280}{arXiv: hep-ph/9301280}].





\bibitem{Giovannini:2001} M.~Giovannini,
Variation of the gauge couplings during inflation,
\href{https://doi.org/10.1103/PhysRevD.64.061301}{Phys. Rev. D \textbf{64}, 061301(R) (2001)}
[\href{https://arxiv.org/abs/astro-ph/0104290}{arXiv: astro-ph/0104290}].
	
\bibitem{Bamba:2004} K.~Bamba and J.~Yokoyama,
Large-scale magnetic fields from inflation in dilaton electromagnetism,
\href{https://doi.org/10.1103/PhysRevD.69.043507}{Phys. Rev. D \textbf{69}, 043507 (2004)}
[\href{https://arxiv.org/abs/astro-ph/0310824}{arXiv: astro-ph/0310824}].
	
\bibitem{Martin:2008} J.~Martin and J.~Yokoyama,
Generation of large scale magnetic fields in single-field inflation, \href{https://doi.org/10.1088/1475-7516/2008/01/025}{J. Cosmol. Astropart. Phys. 01 (2008) 025}
[\href{https://arxiv.org/abs/0711.4307}{arXiv: 0711.4307 [astro-ph]}].
	
\bibitem{Kanno:2009} S.~Kanno, J.~Soda, and M.~Watanabe,
Cosmological magnetic fields from inflation and backreaction, \href{https://doi.org/10.1088/1475-7516/2009/12/009}{J. Cosmol. Astropart. Phys. 12 (2009) 009}
[\href{https://arxiv.org/abs/0908.3509}{arXiv: 0908.3509 [astro-ph.CO]}].
	
\bibitem{Demozzi:2009} V.~Demozzi, V.M.~Mukhanov, and H.~Rubinstein,
Magnetic fields from inflation?,
\href{https://doi.org/10.1088/1475-7516/2009/08/025}{J. Cosmol. Astropart. Phys. 08 (2009) 025}
[\href{https://arxiv.org/abs/0907.1030}{arXiv: 0907.1030 [astro-ph.CO]}].

\bibitem{Maleknejad:2012} A.~Maleknejad, M.M.~Sheikh-Jabbari, and J.~Soda,
Gauge Fields and Inflation,
\href{https://doi.org/10.1016/j.physrep.2013.03.003}{Phys. Rept. \textbf{528}, 161 (2013)}
[\href{https://arxiv.org/abs/1212.2921}{arXiv:1212.2921 [hep-th]}].
	
\bibitem{Ferreira:2013} R.J.Z.~Ferreira, R.K.~Jain, and M.S.~Sloth,
Inflationary magnetogenesis without the strong coupling problem,
\href{https://doi.org/10.1088/1475-7516/2013/10/004}{J. Cosmol. Astropart. Phys. 10 (2013) 004}
[\href{https://arxiv.org/abs/1305.7151}{arXiv: 1305.7151 [astro-ph.CO]}].
	
\bibitem{Ferreira:2014} R.J.Z.~Ferreira, R.K.~Jain, and M.S.~Sloth,
Inflationary magnetogenesis without the strong coupling problem. II. Constraints from CMB anisotropies and B-modes,
\href{https://doi.org/10.1088/1475-7516/2014/06/053}{J. Cosmol. Astropart. Phys. 06 (2014) 053}
[\href{https://arxiv.org/abs/1403.5516}{arXiv: 1403.5516 [astro-ph.CO]}].
	
\bibitem{Vilchinskii:2017} S.~Vilchinskii, O.~Sobol, E.V.~Gorbar, and I.~Rudenok,
Magnetogenesis during inflation and preheating in the Starobinsky model,
\href{https://doi.org/10.1103/PhysRevD.95.083509}{Phys. Rev. D \textbf{95}, 083509 (2017)}
[\href{https://arxiv.org/abs/1702.02774}{arXiv: 1702.02774 [astro-ph.CO]}].
	
\bibitem{Sharma:2017b} R.~Sharma, S.~Jagannathan, T.R.~Seshadri, and K.~Subramanian,
Challenges in inflationary magnetogenesis: Constraints from strong coupling, backreaction, and the Schwinger effect,
\href{https://doi.org/10.1103/PhysRevD.96.083511}{Phys. Rev. D \textbf{96}, 083511 (2017)}
[\href{https://arxiv.org/abs/1708.08119}{arXiv: 1708.08119 [astro-ph.CO]}].
	
\bibitem{Sobol:2018} O.O.~Sobol, E.V.~Gorbar, M.~Kamarpour, and S.I.~Vilchinskii,
Influence of backreaction of electric fields and Schwinger effect on inflationary magnetogenesis,
\href{https://doi.org/10.1103/PhysRevD.98.063534}{Phys. Rev. D \textbf{98}, 063534 (2018)}
[\href{https://arxiv.org/abs/1807.09851}{arXiv: 1807.09851 [hep-ph]}].
	
\bibitem{Talebian:2020}	A.~Talebian, A.~Nassiri-Rad and H.~Firouzjahi,
Revisiting magnetogenesis during inflation,
\href{https://doi.org/10.1103/PhysRevD.102.103508}{Phys. Rev. D \textbf{102}, 103508 (2020)}
[\href{https://arxiv.org/abs/2007.11066}{arXiv: 2007.11066 [gr-qc]}].

\bibitem{Sobol:2020lec} O.O.~Sobol, A.V.~Lysenko, E.V.~Gorbar, and S.I.~Vilchinskii,
Gradient expansion formalism for magnetogenesis in the kinetic coupling model,
\href{https://doi.org/10.1103/PhysRevD.102.123512}{Phys.\ Rev.\ D \textbf{102}, 123512 (2020)}
[\href{http://arxiv.org/abs/2010.13587}{arXiv: 2010.13587 [astro-ph.CO]}].

\bibitem{Sasaki:2022} M.~Sasaki, V.~Vardanyan, and V.~Yingcharoenrat,
Super-horizon resonant magnetogenesis during inflation,
\href{https://doi.org/10.1103/PhysRevD.107.083517}{Phys. Rev. D \textbf{107}, 083517 (2023)}
[\href{http://arxiv.org/abs/2210.07050}{arXiv:2210.07050 [astro-ph.CO]}].



	
\bibitem{Durrer:2011} R.~Durrer, L.~Hollenstein, and R.K.~Jain,
Can slow roll inflation induce relevant helical magnetic fields?,
\href{https://doi.org/10.1088/1475-7516/2011/03/037}{J. Cosmol. Astropart. Phys. 03 (2011) 037}
[\href{http://arxiv.org/abs/1005.5322}{arXiv: 1005.5322 [astro-ph.CO]}].
	
\bibitem{Anber:2006} M.M.~Anber and L.~Sorbo,
N-flationary magnetic fields,
\href{https://doi.org/10.1088/1475-7516/2006/10/018}{J. Cosmol. Astropart. Phys. 10 (2006) 018}
[\href{https://arxiv.org/abs/astro-ph/0606534}{arXiv: astro-ph/0606534}].
	
\bibitem{Anber:2010} M.M.~Anber and L.~Sorbo,
Naturally inflating on steep potentials through electromagnetic dissipation,
\href{https://doi.org/10.1103/PhysRevD.81.043534}{Phys. Rev. D \textbf{81}, 043534 (2010)}
[\href{https://arxiv.org/abs/0908.4089}{arXiv: 0908.4089 [hep-th]}].
	
\bibitem{Barnaby:2012} N.~Barnaby, E.~Pajer, and M.~Peloso,
Gauge field production in axion inflation: consequences for monodromy, non-Gaussianity in the CMB, and gravitational waves at interferometers,
\href{https://doi.org/10.1103/PhysRevD.85.023525}{Phys. Rev. D \textbf{85}, 023525 (2012)}
[\href{https://arxiv.org/abs/1110.3327}{arXiv: 1110.3327 [astro-ph.CO]}].
	
\bibitem{Caprini:2014} C.~Caprini and L.~Sorbo,
Adding helicity to inflationary	magnetogenesis,
\href{https://doi.org/10.1088/1475-7516/2014/10/056}{J. Cosmol. Astropart. Phys. 10 (2014) 056}
[\href{https://arxiv.org/abs/1407.2809}{arXiv: 1407.2809 [astro-ph.CO]}].
	
\bibitem{Anber:2015} M.M.~Anber and E.~Sabancilar,
Hypermagnetic fields and baryon asymmetry from pseudoscalar inflation,
\href{https://doi.org/10.1103/PhysRevD.92.101501}{Phys. Rev. D \textbf{92}, 101501(R) (2015)}
[\href{https://arxiv.org/abs/1507.00744}{arXiv: 1507.00744 [hep-th]}].
	
\bibitem{Ng:2015} K.-W.~Ng, S.-L.~Cheng, and W.~Lee,
Inflationary dilaton-axion magnetogenesis, \href{https://doi.org/10.6122/CJP.20150909}{Chin. J. Phys. \textbf{53}, 110105 (2015)}
[\href{https://arxiv.org/abs/1409.2656}{arXiv:1409.2656 [astro-ph.CO]}].
	
\bibitem{Fujita:2015} T.~Fujita, R.~Namba, Y.~Tada, N.~Takeda, and H.~Tashiro,
Consistent generation of magnetic fields in axion inflation models,
\href{https://doi.org/10.1088/1475-7516/2015/05/054}{J. Cosmol. Astropart. Phys. 05 (2015) 054}
[\href{http://arxiv.org/abs/1503.05802}{arXiv: 1503.05802 [astro-ph.CO]}].
	
\bibitem{Adshead:2015} P.~Adshead, J.T.~Giblin, Jr., T.R.~Scully, and E.I.~Sfakianakis,
Gauge-preheating and the end of axion inflation,
\href{https://doi.org/10.1088/1475-7516/2015/12/034}{J. Cosmol. Astropart. Phys. 12 (2015) 034}
[\href{https://arxiv.org/abs/1502.06506}{arXiv: 1502.06506 [astro-ph.CO]}].
	
\bibitem{Adshead:2016} P.~Adshead, J.T.~Giblin, Jr., T.R.~Scully, and E.I.~Sfakianakis,
Magnetogenesis from axion inflation,
\href{https://doi.org/10.1088/1475-7516/2016/10/039}{J. Cosmol. Astropart. Phys. 10 (2016) 039}
[\href{https://arxiv.org/abs/1606.08474}{arXiv: 1606.08474 [astro-ph.CO]}].
	
\bibitem{Notari:2016} A.~Notari and K.~Tywoniuk,
Dissipative axial inflation,
\href{https://doi.org/10.1088/1475-7516/2016/12/038}{J. Cosmol. Astropart. Phys. 12 (2016) 038}
[\href{http://arxiv.org/abs/1608.06223}{arXiv: 1608.06223 [hep-th]}].
	
\bibitem{Domcke:2018eki} V.~Domcke and K.~Mukaida,
Gauge field and fermion production during axion inflation,
\href{https://doi.org/10.1088/1475-7516/2018/11/020}{J. Cosmol. Astropart. Phys. 11 (2018) 020}
[\href{https://arxiv.org/abs/1806.08769}{arXiv: 1806.08769 [hep-ph]}].
		
\bibitem{Cuissa:2018} J.R.C.~Cuissa and D.G.~Figueroa,
Lattice formulation of axion inflation. Application to preheating,
\href{https://doi.org/10.1088/1475-7516/2019/06/002}{J. Cosmol. Astropart. Phys. 06 (2019) 002}
[\href{https://arxiv.org/abs/1812.03132}{arXiv: 1812.03132 [astro-ph.CO]}].
	
\bibitem{Shtanov:2019} Yu.~Shtanov,
Viable inflationary magnetogenesis with helical coupling,
\href{https://doi.org/10.1088/1475-7516/2019/10/008}{J. Cosmol. Astropart. Phys. 10 (2019) 008}
[\href{https://arxiv.org/abs/1902.05894}{arXiv: 1902.05894 [astro-ph.CO]}].
		
\bibitem{Sobol:2019} O.O.~Sobol, E.V.~Gorbar, and S.I.~Vilchinskii,
Backreaction of electromagnetic fields and the Schwinger effect in pseudoscalar inflation magnetogenesis,
\href{https://doi.org/10.1103/PhysRevD.100.063523}{Phys. Rev. D \textbf{100}, 063523 (2019)}
[\href{http://arxiv.org/abs/1907.10443}{arXiv: 1907.10443 [astro-ph.CO]}].

\bibitem{Domcke:2020} V.~Domcke, V.~Guidetti, Y.~Welling, and A.~Westphal,
Resonant backreaction in axion inflation,
\href{https://doi.org/10.1088/1475-7516/2020/09/009}{J. Cosmol. Astropart. Phys. 09 (2020) 009}
[\href{https://arxiv.org/abs/2002.02952}{arXiv: 2002.02952 [astro-ph.CO]}].

\bibitem{Caravano:2021} A.~Caravano, E.~Komatsu, K.D.~Lozanov, and J.~Weller,
Lattice simulations of Abelian gauge fields coupled to axions during inflation,
\href{https://doi.org/10.1103/PhysRevD.105.123530}{Phys. Rev. D \textbf{105}, 123530 (2022)}
[\href{https://arxiv.org/abs/2110.10695}{arXiv:2110.10695 [astro-ph.CO]}].

\bibitem{Gorbar:2021rlt} E.V.~Gorbar, K.~Schmitz, O.O.~Sobol, and S.I. Vilchinskii,
Gauge-field production during axion inflation in the gradient-expansion formalism,
\href{https://doi.org/10.1103/PhysRevD.104.123504}{Phys. Rev. D \textbf{104}, 123504 (2021)}
[\href{https://arxiv.org/abs/2109.01651}{arXiv:2109.01651 [hep-ph]}].

\bibitem{Gorbar:2022} E.V.~Gorbar, K.~Schmitz, O.O.~Sobol, and S.I.~Vilchinskii,
Hypermagnetogenesis from axion inflation: Model-independent estimates,
\href{https://doi.org/10.1103/PhysRevD.105.043530}{Phys. Rev. D \textbf{105}, 043530 (2022)}
[\href{https://arxiv.org/abs/2111.04712}{arXiv:2111.04712 [hep-ph]}].

\bibitem{Adshead:2018doq} P.~Adshead, J.T.~Giblin, Jr., and Z.J.~Weiner,
Gravitational waves from gauge preheating,
\href{https://doi.org/10.1103/PhysRevD.98.043525}{Phys. Rev. D \textbf{98}, 043525 (2018)}
[\href{https://arxiv.org/abs/1805.04550}{arXiv:1805.04550 [astro-ph.CO]}].

\bibitem{Adshead:2019igv} P.~Adshead, J.T.~Giblin, Jr., M.~Pieroni, and Z.J.~Weiner,
Constraining Axion Inflation with Gravitational Waves across 29 Decades in Frequency,
\href{https://doi.org/10.1103/PhysRevLett.124.171301}{Phys. Rev. Lett. \textbf{124}, 171301 (2020)}
[\href{https://arxiv.org/abs/1909.12843}{arXiv:1909.12843 [astro-ph.CO]}].

\bibitem{Adshead:2019lbr} P.~Adshead, J.T.~Giblin, Jr., M.~Pieroni, and Z.J.~Weiner,
Constraining axion inflation with gravitational waves from preheating,
\href{https://doi.org/10.1103/PhysRevD.101.083534}{Phys. Rev. D \textbf{101}, 083534 (2020)}
[\href{https://arxiv.org/abs/1909.12842}{arXiv:1909.12842 [astro-ph.CO]}].

\bibitem{Caravano:2022}
A.~Caravano, E.~Komatsu, K.D.~Lozanov, and J.~Weller,
Lattice Simulations of Axion-U(1) Inflation,
\href{https://doi.org/10.1103/PhysRevD.108.043504}{Phys. Rev. D \textbf{108}, 043504 (2023)}
[\href{https://arxiv.org/abs/2204.12874}{arXiv:2204.12874 [astro-ph.CO]}].

\bibitem{Bastero-Gil:2022}
M.~Bastero-Gil and A.T.~Manso,
Parity violating gravitational waves at the end of inflation,
\href{https://doi.org/10.1088/1475-7516/2023/08/001}{J. Cosmol. Astropart. Phys. 08 (2023) 001}
[\href{https://arxiv.org/abs/2209.15572}{arXiv:2209.15572 [gr-qc]}].

\bibitem{Fujita:2022}
T.~Fujita, K.~Mukaida and Y.~Tada,
Stochastic formalism for U(1) gauge fields in axion inflation,
\href{https://doi.org/10.1088/1475-7516/2022/12/026}{J. Cosmol. Astropart. Phys. 12 (2022) 026}
[\href{https://arxiv.org/abs/2206.12218}{arXiv:2206.12218 [astro-ph.CO]}].

	
\bibitem{Savchenko:2018} O.~Savchenko and Yu.~Shtanov,
Magnetogenesis by non-minimal coupling to gravity in the Starobinsky inflationary model,
\href{https://doi.org/10.1088/1475-7516/2018/10/040}{J. Cosmol. Astropart. Phys. 10 (2018) 040}
[\href{https://arxiv.org/abs/1808.06193}{arXiv:1808.06193 [astro-ph.CO]}].
	
\bibitem{Sobol:2021} O.O.~Sobol, E.V.~Gorbar, O.M.~Teslyk, and S.I.~Vilchinskii,
Generation of an electromagnetic field nonminimally coupled to gravity during Higgs inflation,
\href{https://doi.org/10.1103/PhysRevD.104.043509}{Phys. Rev. D \textbf{104}, 043509 (2021)}
[\href{https://arxiv.org/abs/2104.14400}{arXiv:2104.14400 [gr-qc]}].

\bibitem{Maity:2021} D.~Maity, S.~Pal, and T.~Paul,
Effective theory of inflationary magnetogenesis and constraints on reheating,
\href{https://doi.org/10.1088/1475-7516/2021/05/045}{J. Cosmol. Astropart. Phys. 05 (2021) 045}
\href{https://arxiv.org/abs/2103.02411}{arXiv: 2103.02411 [hep-th]}.

\bibitem{Durrer:2022emo} R.~Durrer, O.~Sobol, and S.~Vilchinskii,
Magnetogenesis in Higgs-Starobinsky inflation,
\href{https://doi.org/10.1103/PhysRevD.106.123520}{Phys. Rev. D \textbf{106}, 123520 (2022)}
[\href{http://arxiv.org/abs/2207.05030}{arXiv:2207.05030 [gr-qc]}].

\bibitem{Bamba:2008} K.~Bamba and S.D.~Odintsov,
Inflation and late-time cosmic acceleration in non-minimal Maxwell-$F(R)$ gravity and the generation of large-scale magnetic fields,
\href{https://doi.org/10.1088/1475-7516/2008/04/024}{J. Cosmol. Astropart. Phys. 04 (2008) 024}
[\href{https://arxiv.org/abs/0801.0954}{arXiv: 0801.0954 [astro-ph]}].
	
\bibitem{Bamba:2020} K.~Bamba, E.~Elizalde, S.D.~Odintsov, and T.~Paul,
Inflationary magnetogenesis with reheating phase from higher curvature coupling,
\href{https://doi.org/10.1088/1475-7516/2021/04/009}{J. Cosmol. Astropart. Phys. 04 (2021) 009}
[\href{https://arxiv.org/abs/2012.12742}{arXiv: 2012.12742 [gr-qc]}].

\bibitem{Cecchini:2023} 
C.~Cecchini and M.~Rinaldi,
Inflationary helical magnetic fields with a sawtooth coupling,
\href{https://doi.org/10.1016/j.dark.2023.101212}{Phys. Dark Univ. \textbf{40}, 101212 (2023)}
[\href{https://arxiv.org/abs/2301.07699}{arXiv:2301.07699 [astro-ph.CO]}].



\bibitem{Sugiyama:2012}
N.S.~Sugiyama, E.~Komatsu, and T.~Futamase, 
$\delta N$ formalism,
\href{https://doi.org/10.1103/PhysRevD.87.023530}{Phys. Rev. D \textbf{87}, 023530 (2013)}
[\href{https://arxiv.org/abs/1208.1073}{arXiv:1208.1073 [gr-qc]}].





\bibitem{Fischetti:1979} M.V.~Fischetti, J.B.~Hartle, and B.L.~Hu,
Quantum Effects in the Early Universe. 1. Influence of Trace Anomalies on Homogeneous, Isotropic, Classical Geometries,
\href{https://doi.org/10.1103/PhysRevD.20.1757}{Phys. Rev. D \textbf{20}, 1757 (1979)}.


\bibitem{Bezrukov:2008} F.L.~Bezrukov and M.~Shaposhnikov,
The Standard Model Higgs boson as the inflaton,
\href{https://doi.org/10.1016/j.physletb.2007.11.072}{Phys. Lett. B \textbf{659}, 703 (2008)}
[\href{https://arxiv.org/abs/0710.3755}{arXiv: 0710.3755 [hep-th]}].
	
\bibitem{Bauer:2008} F.~Bauer and D.A.~Demir,
Inflation with non-minimal coupling: Metric versus Palatini formulations,
\href{https://doi.org/10.1016/j.physletb.2008.06.014}{Phys. Lett. B \textbf{665}, 222 (2008)}
[\href{https://arxiv.org/abs/0803.2664}{arXiv: 0803.2664 [hep-ph]}].

\bibitem{Ema:2017} Y.~Ema,
Higgs Scalaron Mixed Inflation,
\href{https://doi.org/10.1016/j.physletb.2017.04.060}{Phys. Lett. B \textbf{770}, 403 (2017)}
[\href{https://arxiv.org/abs/1701.07665}{arXiv:1701.07665 [hep-ph]}].
    
\bibitem{Starobinsky:2018} M.~He, A.~A.~Starobinsky, and J.~Yokoyama,
Inflation in the mixed Higgs-$R^2$ model,
\href{https://doi.org/10.1088/1475-7516/2018/05/064}{J. Cosmol. Astropart. Phys. 05 (2018) 064}
[\href{https://arxiv.org/abs/1804.00409}{arXiv:1804.00409 [astro-ph.CO]}].

\bibitem{Gorbunov:2018} D.~Gorbunov and A.~Tokareva,
Scalaron the healer: removing the strong-coupling in the Higgs- and Higgs-dilaton inflations,
\href{https://doi.org/10.1016/j.physletb.2018.11.015}{Phys. Lett. B \textbf{788}, 37 (2019)}
[\href{https://arxiv.org/abs/1807.02392}{arXiv:1807.02392 [hep-ph]}].

\bibitem{Bunch:1978} T.S.~Bunch and P.C.W.~Davies,
Quantum field theory in de Sitter space: Renormalization by point splitting,
\href{https://doi.org/10.1098/rspa.1978.0060}{Proc. R. Soc. A \textbf{360}, 117 (1978)}.

\bibitem{Harko:2010}
T.~Harko and F.S.N.~Lobo,
f(R,$L_{m}$) gravity,
\href{https://doi.org/10.1140/epjc/s10052-010-1467-3}{Eur. Phys. J. C \textbf{70}, 373 (2010)}
[\href{https://arxiv.org/abs/1008.4193}{arXiv:1008.4193 [gr-qc]}].

\bibitem{BeltranJimenez:2017} J.~Beltr\'{a}n Jim\'{e}nez, L.~Heisenberg, G.J.~Olmo, and D.~Rubiera-Garcia,
Born-Infeld inspired modifications of gravity,
\href{https://doi.org/10.1016/j.physrep.2017.11.001}{Phys. Rept. \textbf{727}, 1 (2018)}
[\href{https://arxiv.org/abs/1704.03351}{arXiv:1704.03351 [gr-qc]}].

\bibitem{Peloso:2022} M.~Peloso and L.~Sorbo,
Instability in axion inflation with strong backreaction from gauge modes,
\href{https://doi.org/10.1088/1475-7516/2023/01/038}{J. Cosmol. Astropart. Phys. 01 (2023) 038}
[\href{https://arxiv.org/abs/2209.08131}{arXiv:2209.08131 [astro-ph.CO]}].

\bibitem{AbramowitzStegun} M.~Abramowitz and I.A.~Stegun,
\textit{Handbook of mathematical functions with formulas, graphs, and mathematical tables}
(U.S. Government Printing Office, Washington, D.C., 1964).




	





	

\end{thebibliography}
\end{document}